\documentclass[sts]{imsart}

\RequirePackage{amsthm,amsmath,amsfonts,amssymb}
\RequirePackage[authoryear]{natbib}
\RequirePackage{graphicx}
\RequirePackage{float}
\RequirePackage{bm}
\startlocaldefs

\newcommand{\xb}{\mathbf{x}}
\newcommand{\yb}{\mathbf{y}}

\newcommand{\bb}{\bm{b}}

\newcommand{\Db}{\mathbf{D}}

\newcommand{\Mb}{\mathbf{M}}

\newcommand{\Ub}{\mathbf{U}}
\newcommand{\Vb}{\mathbf{V}}

\newcommand{\Xb}{\mathbf{X}}

\newcommand{\bbR}{\mathbb{R}}

\newcommand{\bmu}{\bm{\mu}}
\newcommand{\Sigmab}{\bm{\Sigma}}

\newcommand{\0}{{\mathbf{0}}}
\newcommand{\1}{{\mathbf{1}}}

\endlocaldefs

\begin{document}

\begin{frontmatter}
\title{New Perspectives on Centering}
\runtitle{New Perspectives on Centering}

\begin{aug}
\author[A]{\fnms{Jack B.} \snm{Prothero}\ead[label=e1]{jackb37@live.unc.edu}},
\author[B]{\fnms{Jan} \snm{Hannig}\ead[label=e2]{jan.hannig@unc.edu}}
\and
\author[C]{\fnms{J.S.} \snm{Marron}\ead[label=e3]{MARRON@unc.edu}}
\address[A]{Jack B. Prothero is Ph.D. candidate, Department of Statistics and Operations Research, University of North Carolina Chapel Hill, USA, \printead{e1}.}
\address[B]{Jan Hannig is Professor, Department of Statistics and Operations Research, University of North Carolina Chapel Hill, USA, \printead{e2}.}
\address[C]{J.S. Marron is Amos Hawley Distinguished Professor, Department of Statistics and Operations Research, University of North Carolina Chapel Hill, USA, \printead{e3}.}
\end{aug}

\begin{abstract}
Data matrix centering is an ever-present yet under-examined aspect of data analysis. Functional data analysis (FDA) often operates with a default of centering such that the vectors in one dimension have mean zero. We find that centering along the other dimension identifies a novel useful mode of variation beyond those familiar in FDA. We explore ambiguities in both matrix orientation and nomenclature. Differences between centerings and their potential interaction can be easily misunderstood. We propose a unified framework and new terminology for centering operations. We clearly demonstrate the intuition behind and consequences of each centering choice with informative graphics. We also propose a new direction energy hypothesis test as part of a series of diagnostics for determining which choice of centering is best for a data set. We explore the application of these diagnostics in several FDA settings.
\end{abstract}


\end{frontmatter}

\section{Introduction}

Many data processing pipelines involve transformations such as \textit{centering}: the subtraction of the mean of a set of values resulting in the transformed data having 0 mean. Despite the pervasiveness of such transformations, there are surprising misunderstandings concerning their meaning and implications. In this paper we present a survey of the effects of different forms of centering on a data matrix and the consequences of those effects within widely-used data analysis methods. We first seek to disambiguate the terminology used to discuss centering colloquially by putting forth a carefully-considered nomenclature framework. With a unified lexical understanding we discuss the geometric effects of each centering in all relevant vector spaces. We find overall that new and more complete data insights are available via a new mode of variation derived from non-standard centering. The case studies and hypothesis tests presented in this paper provide a blueprint for how to determine which centering to ultimately opt for in new analyses.

In Functional Data Analysis (FDA), data are commonly organized in a $d\times n$ matrix with one of rows or columns considered as \textit{curves}. In this paper, we follow the convention in \cite{ooda} of columns representing those curves. We refer to each $d$-dimensional column vector as a \textit{data object} (i.e. experimental unit, data point, observation, case). This terminology appropriately reflects the full generality of the kinds of data collected and stored in matrix form in modern settings. We refer to each $n$-dimensional row vector as a \textit{trait} (i.e. feature, variable). While this term is non-standard, it avoids potential ambiguity in using the more popular term "feature." In some areas of data science, "feature vector" refers to what are called data objects here. Because vectors along both dimensions of data matrices are critical to our discussion, we need an appropriately distinct name for each. As one of the main goals of this paper is disambiguating the terminology surrounding centering, we prefer the term "trait vector" to represent the vectors in the dimension opposite the data objects.

Some researchers and software packages opt for the transpose of our convention: using columns as traits and rows as data objects. In fact, the legacy of structural limitations of data analysis software reverberates through our choices of matrix orientation to this day. Many tools placed stricter limits on the number of columns a data table could have, mirroring mathematical preferences for "long and skinny" matrices. Most fields during this time period analyzed data with many more objects than traits, so data was typically entered and stored such that objects were rows and traits were columns. Bioinformatics and related fields were in the opposite position and often collected data on a very large number of traits from a relatively small group of objects. This led to data matrices being stored with the opposite orientation: data objects as columns and traits as rows. We follow the bioinformatics convention here. An agreement as to whether rows or columns are data objects is important to facilitate discussion of data analysis between fields. As we'll see shortly, ambiguity in matrix orientation choice has an acutely confounding effect when discussing centering choices.

A time-honored, broadly-used tool in FDA is principal component analysis (PCA). PCA decomposes the data into \textit{modes of variation} about the mean of the data objects. These modes can be calculated from an eigenanalysis of the covariance matrix of the data. To construct the covariance matrix one must first center the data matrix such that the data objects have a mean vector of 0. While this choice of centering is very natural, it is unclear whether it should be called "column centering" or "row centering" regardless of matrix orientation convention. In our convention one might first consider this a vector operation and call it "column centering". However, the operation is equivalent to finding the mean value of each trait row vector and subtracting it from each of that trait row vector's entries. From this perspective the operation could be called "row centering" as the entries of each row have mean zero after the operation. In the other matrix orientation convention, the same could be said of "column centering." We propose new terminology specifically aimed at avoiding this sort of ambiguity. As this translation of the data objects in $\bbR^d$ (data object space) such that they are centered at the origin is an important effect of this centering operation, we will refer to this operation as \textit{object centering} a data matrix. Referring to the intended target of the centering (object vs trait) as opposed to the matrix dimension (column vs row) clarifies the intended meaning while also unifying terminology regardless of choice of matrix orientation. 

Including object centering the matrix, there are in fact four total centerings available besides leaving the matrix uncentered. 
\begin{itemize}
\item \textit{Trait centering} is the dual operation to object centering. From a vector point of view, the trait vectors are translated in $\bbR^n$ such that their mean vector is at the origin. As a result, the entries of each individual data object have a mean of 0.

\item \textit{Grand mean centering} finds the mean of all entries of the data matrix (the \textit{grand mean}) and subtracts that value from each entry.

\item \textit{Double centering} is the result of performing object centering followed by trait centering (or vice versa, the operations commute) on the matrix. The resulting matrix has all the properties of both object-centered matrices and trait-centered matrices.
\end{itemize}

\cite{jsmsvdcenter} examine the effects of each centering on the quality of low-rank matrix approximations. Here our purpose is more focused on the interpretability and insights from the data gained or lost by using these different forms of centering in statistical analyses. Notably, that manuscript opts for the ambiguous convention of referring to different centerings according to matrix dimension ("row" and "column") rather than according to the goal of the centering operation.  We submit that our nomenclature allows for clearer explorations of these kinds of topics.

\begin{figure}[H]
\centering
\includegraphics[width=\textwidth]{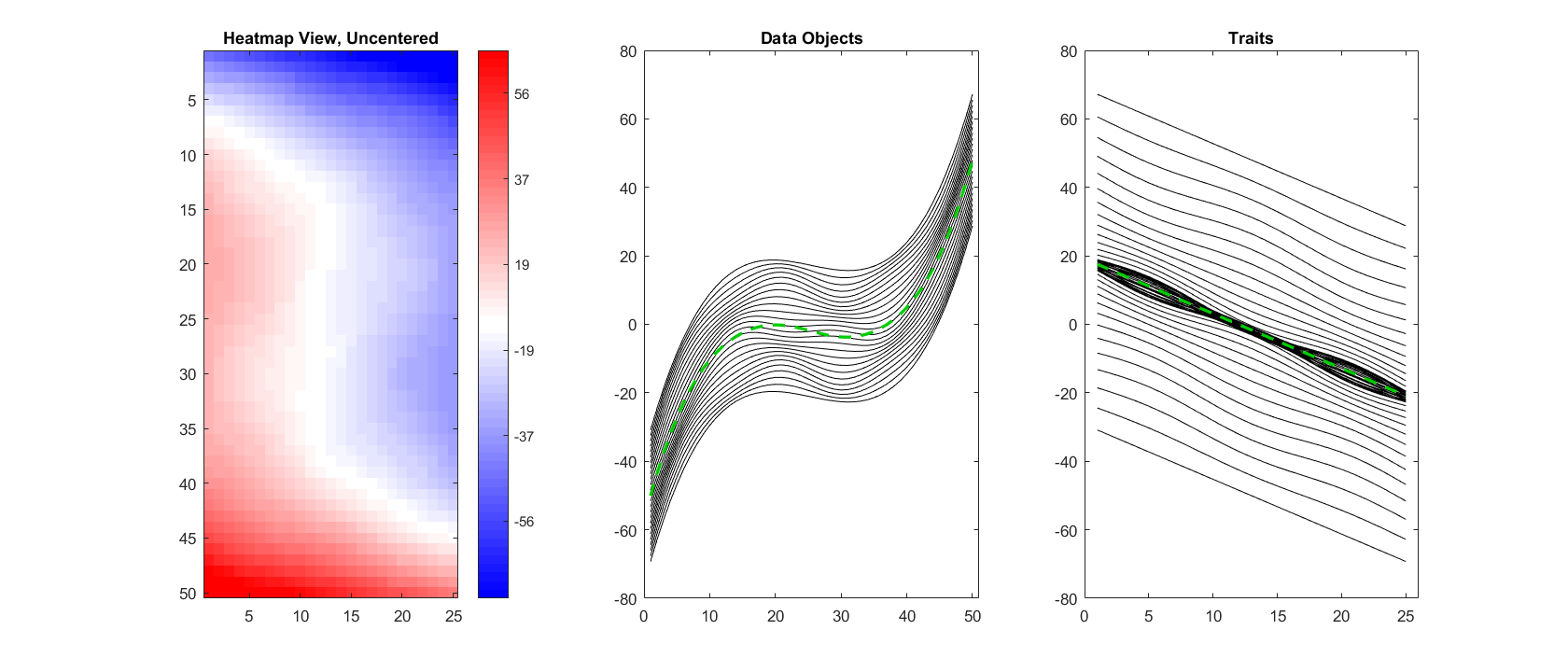}
\caption{Heatmap (left) and functional data (center, right) views of synthetic data example. Heatmap shows a clear undulating pattern. Data object functions are distorted cubic functions and trait functions are distorted linear functions. Green dashed lines are mean curves.}\label{fig:rawToy}
\end{figure}
\begin{figure}[H]
\centering
\includegraphics[width=\textwidth]{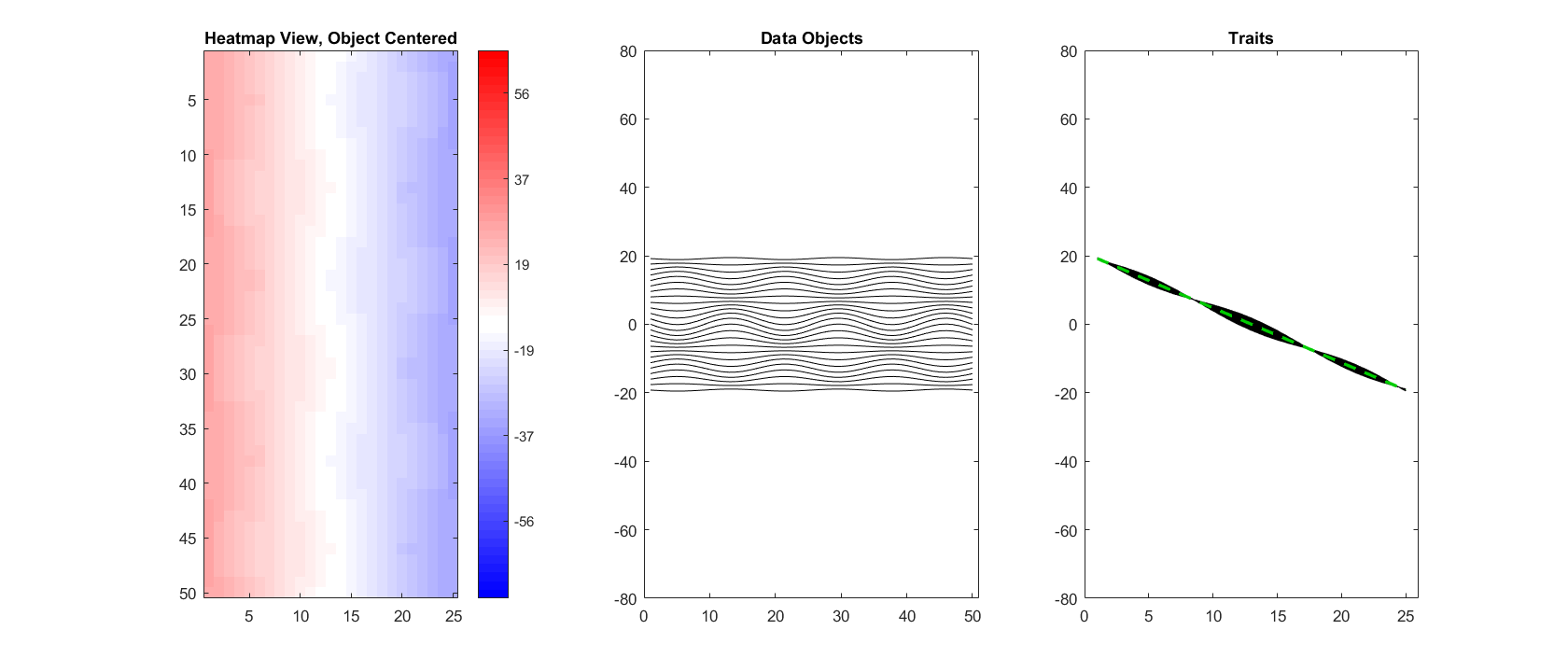}
\caption{Heatmap and functional data view of object-centered synthetic data example. Cubic effect in data objects is removed and linear effect now dominates. Columns are now shifted and scaled low-amplitude sine waves.}\label{fig:objectToy}
\end{figure}
\begin{figure}[H]
\centering
\includegraphics[width=\textwidth]{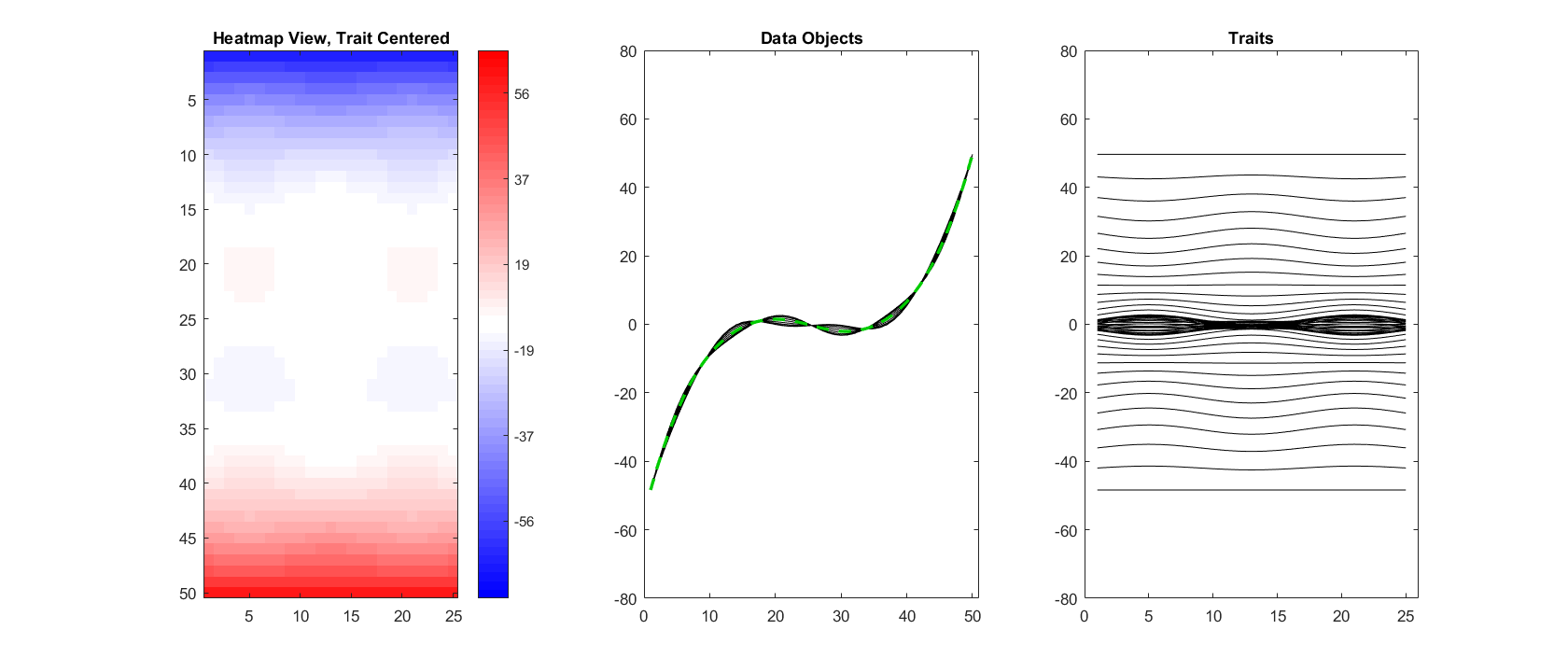}
\caption{Heatmap and functional data view of trait-centered synthetic data example. Linear effect in traits is removed to reveal another small-scale wave effect. Heatmap generally driven by cubic structure in data objects (columns) with faint additional patterns due to underlying sine waves.}\label{fig:featureToy}
\end{figure}
\begin{figure}[H]
\centering
\includegraphics[width=\textwidth]{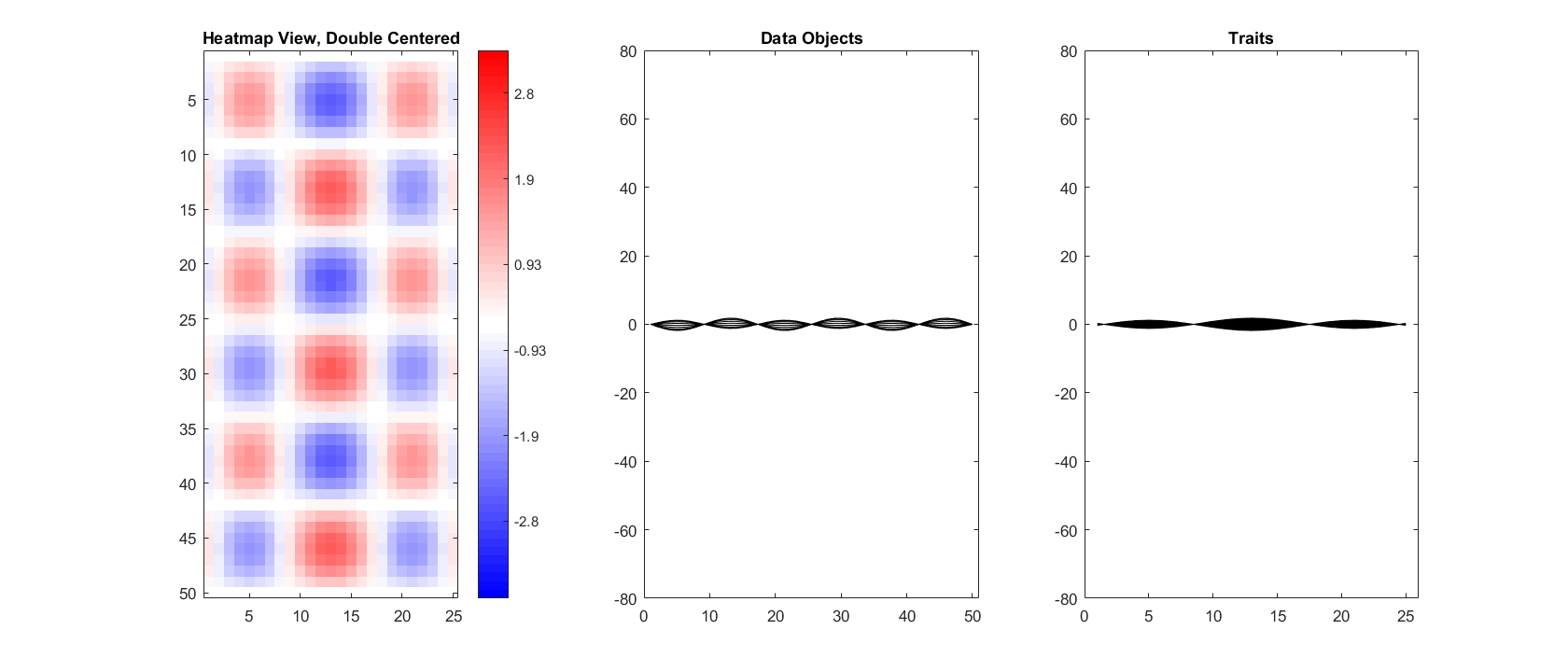}
\caption{Heatmap and functional data view of double-centered synthetic data example. Curve bundles are clear sine waves along either dimension. Heatmap shows a clear planar wave pattern due to the outer product of the sine waves along each dimension. Scale of the heatmap color bar is changed to emphasize this subtle but meaningful effect.}\label{fig:doubleToy}
\end{figure}

Figures~\ref{fig:rawToy}-\ref{fig:doubleToy} visually explore the centerings listed above through different points of view of a common synthetic data set. The synthetic data matrix is $50\times 25$; we display its contents in Figure~\ref{fig:rawToy}. The left panel shows a heatmap view of the data matrix. In a heatmap view, the numerical value of each entry is encoded as a color, with hue indicating a positive (blue) or negative (red) entry and saturation indicating magnitude. The heatmap reveals strong patterns across both the traits and the data objects. We alternatively display these patterns with functional data views of both the data objects and the traits in the center and right panels respectively. The data objects (heatmap columns, center panel curves) are a bundle of distorted and vertically shifted cubic functions with a cubic function mean (center panel green dashed line). The traits (heatmap rows, right panel curves) are a bundle of distorted linear functions with a linear function mean (right panel green dashed line). Curve height in the center \& right panels corresponds with pixel color in the heatmap.
In subsequent views of this data we will perform three of the four centerings on the original matrix and examine the changes in the visual patterns of the data matrix. Each view will substantially and uniquely alter which aspects of the data are prominent and which are hidden. We omit grand mean centering as it amounts to a simple modification of the heatmap colors and a vertical shift of the curves in the functional data plots.

We first perform object centering, with results shown in Figure~\ref{fig:objectToy}. The center panel now shows a series of vertical shifts and amplitude scalings of a sine wave as the cubic structure was removed with the object mean (green dashed line in center panel of Figure~\ref{fig:rawToy}). The right panel shows a strong vertical shift of each of the trait curves, bringing them together around their linear function mean. The heatmap in the left panel is now dominated by the linear effect in each row. We have kept the heatmap color saturation scale and vertical axes in the center \& right panel identical to those in Figure~\ref{fig:rawToy} for effective comparison.

Next we examine the effects of trait centering. As this is the dual operation to object centering, Figure~\ref{fig:featureToy} displays effects dual to those from Figure~\ref{fig:objectToy}. The right panel now shows a series of vertically shifted waves as the sloped linear structure was removed with the trait mean. The center panel shows a strong vertical shift of each of the data object curves, bringing them together around their cubic function mean. The heatmap in the left panel is now dominated by the cubic effect in each column, with some columns containing small, higher-frequency oscillations reflecting the sine wave distortion that was hard to see in Figure~\ref{fig:rawToy}. Once again, this figure uses the same scalings as those in Figure~\ref{fig:rawToy}.

Finally, we perform double centering, which will remove both the cubic function mean among the data objects as well as the linear function mean among the traits. In Figure~\ref{fig:doubleToy}, the residual curves along both dimensions are pure sine waves, and the resulting heatmap in the left panel shows a very clear planar wave pattern.  This underlying mode of variation was obscured in Figure~\ref{fig:rawToy} by the mean effects along either dimension. In this figure the curve plots use the same vertical axes as Figures~\ref{fig:rawToy}-\ref{fig:featureToy} but the color saturation scale of the heatmap is adjusted because the pattern would appear too faint otherwise.

The fundamental point is that each form of centering leads to a substantially different interpretation of the prominent traits of the data matrix. The visual impression of the raw data differs greatly from each of the centered versions. The distinct and interesting pattern that remains in the double-centered data is largely hidden in views containing either the object mean or the trait mean. An important premise of this paper is that paying more attention to this phenomenon can lead to improved insights from exploratory analysis. In particular, we propose a new, insightful mode of variation based on the trait mean for FDA decompositions.

In Section~\ref{sec:fda}, we will analyze mortality and genomic data sets under multiple centering regimes to demonstrate the value of exploring non-standard centerings. In both cases we find enhanced visual interpretability after additional centering operations. In Section~\ref{sec:formalism} we mathematically investigate the geometry of different forms of centering in the dual data object and trait spaces. Combining insights from both of these analyses, in Section~\ref{sec:test} we develop a novel statistical test which determines whether a significant mean effect is lurking as a substantial portion of a mode of variation. Finally, in Section~\ref{sec:pls}, we examine how these lessons on centering can be applied in a multi-block data integration context.

\section{FDA Case Studies} \label{sec:fda}

In FDA, the data objects are typically vectors representing digitized curves. As with other kinds of data objects we're interested in how the objects vary in the space they occupy. We can use traditional tools like PCA and singular value decomposition (SVD) to discover informative modes of variation in the data, and then use the functional interpretation of the data in question to produce insightful visualization of those modes of variation. In the following subsections we present functional data analyses of two data sets: a collection of mortality rates in Spain during the 20th century and a cohort of base-pair level RNAseq observations. In both cases we'll examine the effects that different centering choices have on the visual interpretation of the analysis.

\subsection{Spanish Mortality}

\cite{ooda} consider a data matrix containing mortality rates (proportion of the population of a given age that died in a year) of Spanish males from 1908 to 2002. We are interested in how mortality rates, as a function of age from 0 (birth) to 98, changed over this time span. Hence, we will treat each year as a data object (column) and the mortality rates of each age as a trait (row). We first conduct a classical FDA based on object centering. We then compare those results to a naive uncentered SVD and a double centered FDA.

\begin{figure}[H]
\centering
\includegraphics[scale=0.5]{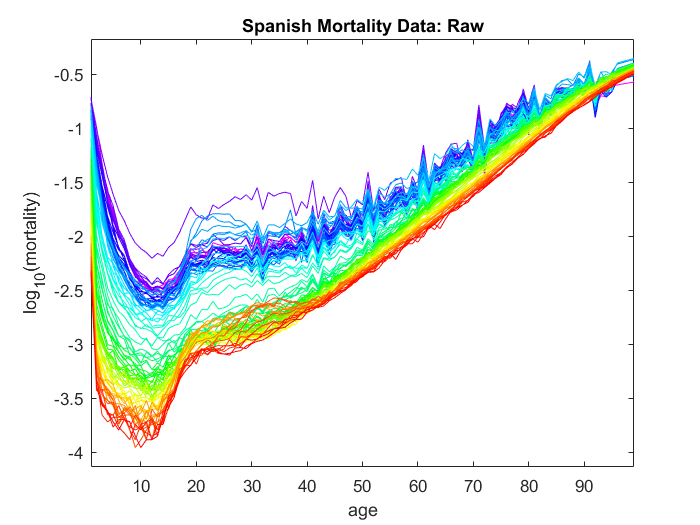}
\caption{Data curve view of the $\log_{10}$ Spanish mortality data. }\label{fig:rawMort}
\end{figure}

Figure~\ref{fig:rawMort} displays the data curves for the Spanish mortality data. Each curve represents a year of data, and the points along each curve encode the mortality rates for each age in that particular year. The curve colors represent chronology, with earlier years displayed in cooler colors and later years displayed in warmer colors. Each entry was adjusted by a $\log_{10}$ transformation because mortality rates tend to vary across several orders of magnitude.

Prominent details include higher mortality rates for newborns and the elderly as well as overall improvement in mortality rate over the course of the 20th century. We observe systematic spikes every 10 years, reflecting strong decadal rounding in death records for older men in the earlier half of the century. 

\begin{figure}[H]
\centering
\includegraphics[scale=0.32]{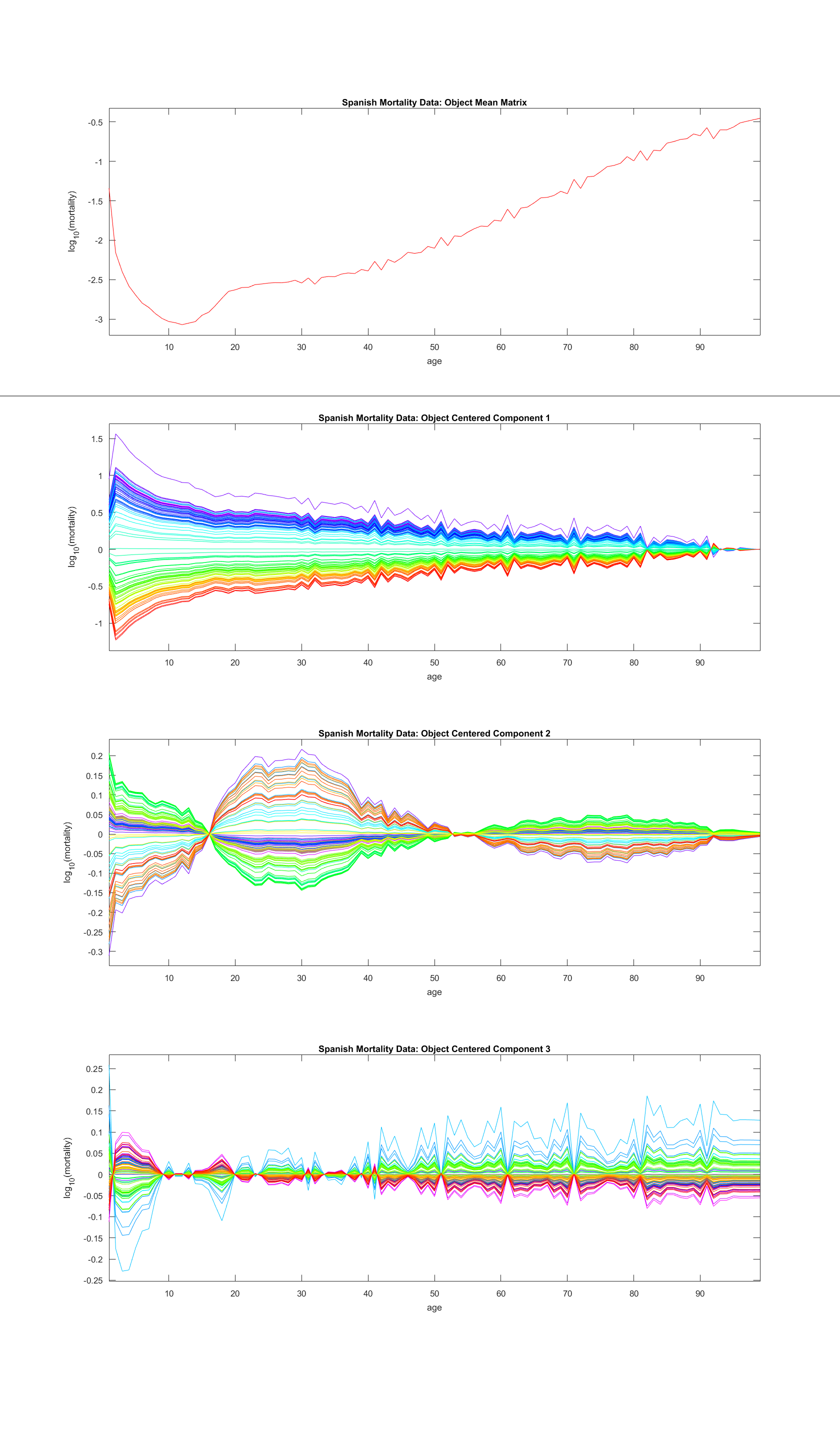}
\caption{Data curve view of the mean and first three principal modes of variation for the Spanish mortality data. Component 1 (second panel) shows overall improvement over time and component 2 (third panel) shows differences between young/middle-age adults and children/the elderly. Note that for each year adding the corresponding curves from each plot together results in an approximation of the original data curves in Figure~\ref{fig:rawMort}.}\label{fig:mortobjectcurves}
\end{figure}

We conduct a conventional FDA to find interesting modes of variation in mortality rates over the course of the 20th century. Insights into these modes of variation come from considering both loadings and scores of a PCA. Figure~\ref{fig:mortobjectcurves} shows the loadings vectors as curves scaled by the scores. The top panel shows the object mean curve as a function of age, and subsequent panels show additional modes of variation about that mean. The second panel (first mode of variation) shows an overall decrease in mortality rate over time which benefitted younger individuals more strongly. The year 1918 is visually distinct at the top of the plot due to the global flu pandemic that year. The third panel (second mode of variation) shows a contrast in mortality rate trends between 18-49 year olds and the rest of the population. This reflects three bursts in mortality for this age group, including the flu pandemic, the Spanish Civil War, and automobile fatalities. Throughout the first and third modes of variation there are remnants of "age-rounding" due to imprecise records. This manifests visually as a repeating pattern over time of length 10.

\begin{figure}[H]
\centering
\includegraphics[width=\textwidth]{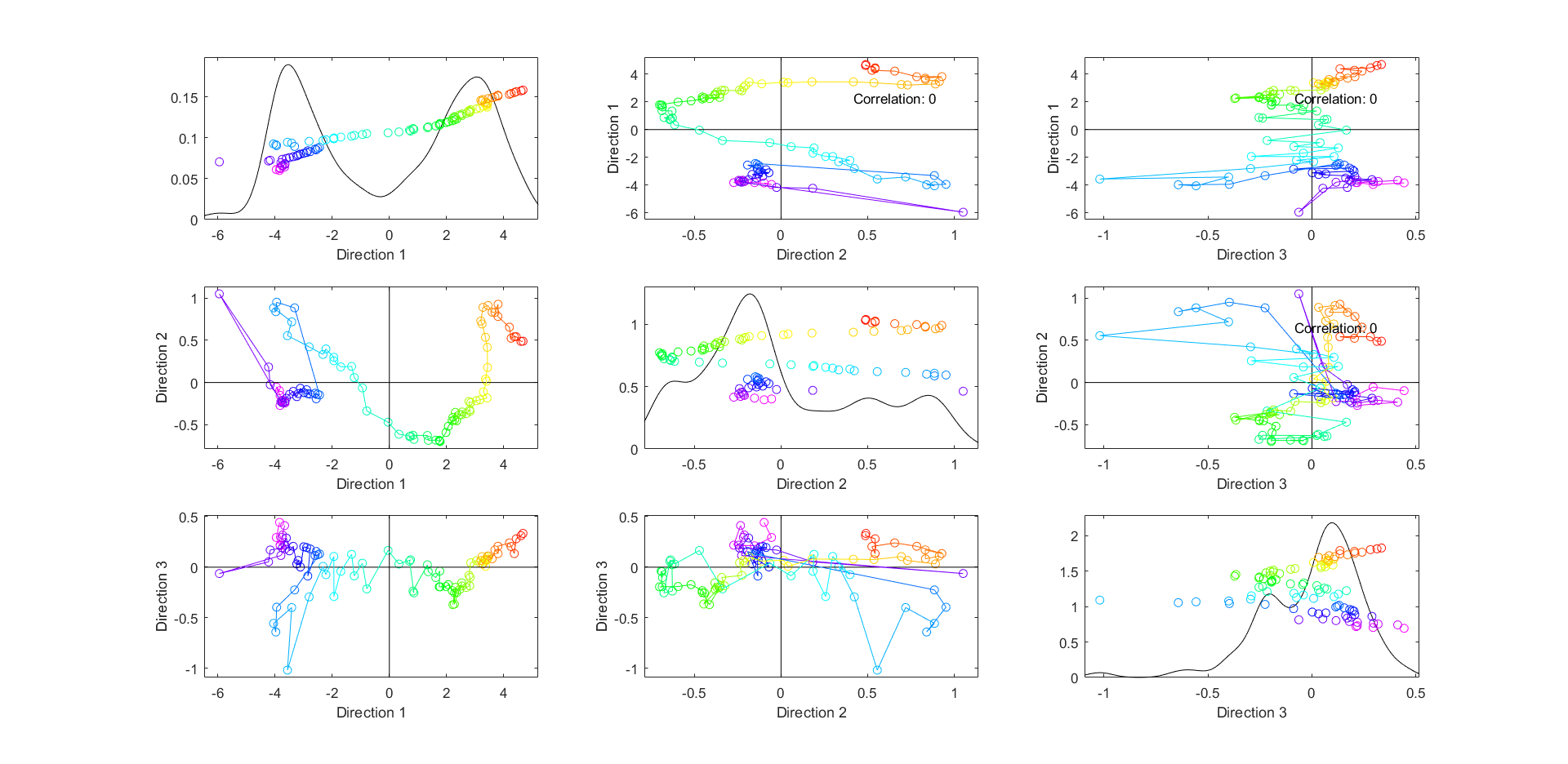}
\caption{Scatter plot view of Spanish mortality data. Most explainable and interpretable trends appear in the two-way plot of components 1 and 2. Due to the object centering performed for conventional FDA, all 2D plots display 0 correlation.}\label{fig:mortobjectscatter}
\end{figure}

While Figure~\ref{fig:mortobjectcurves} explores the modes of variation of the data via loadings, Figure~\ref{fig:mortobjectscatter} explores relationships between the data objects by looking at scatter plots of projections of the data onto score vectors. We generate one-dimensional views of each score vector with score value on the horizontal axis and chronology on the vertical axis overlaid with a smooth histogram. We also plot two-dimensional views showing projections onto the two-dimensional planes generated by each pair of score vectors. These are all organized into a matrix of plots with 1D views on the diagonal and corresponding 2D views in respective off-diagonal slots. The year-based coloring in each plot is consistent with other views. In the 2D scatter plots, we connect the dots in chronological order. The 2D plot between components 1 and 2 shows many of the trends discussed previously. We can track overall improvement over time with obstacles to that improvement arising in the early 20th century (small cluster of blue points in the bottom right) and late 20th century (cluster of orange points in the top right). Notably the correlation in each 2D scatterplot is zero. As we will show in greater detail in Section~\ref{sec:formalism}, this is a consequence of the object centering operation that takes place as the first step of a conventional FDA.

The above PCA can be viewed either as an eigenanalysis of a covariance matrix or as an SVD of the data matrix after it has been object-centered. The right-singular vectors from the SVD are the score vectors and the left-singular vectors are the loadings vectors in our matrix orientation convention. The SVD formulation suggests potential use of other centerings. We could examine left and right singular vectors for an uncentered version of the matrix or a differently-centered version of the matrix. In such cases, the interpretation of the decomposition into modes of variation will typically change substantially. 

\begin{figure}[H]
\centering
\includegraphics[scale=0.32]{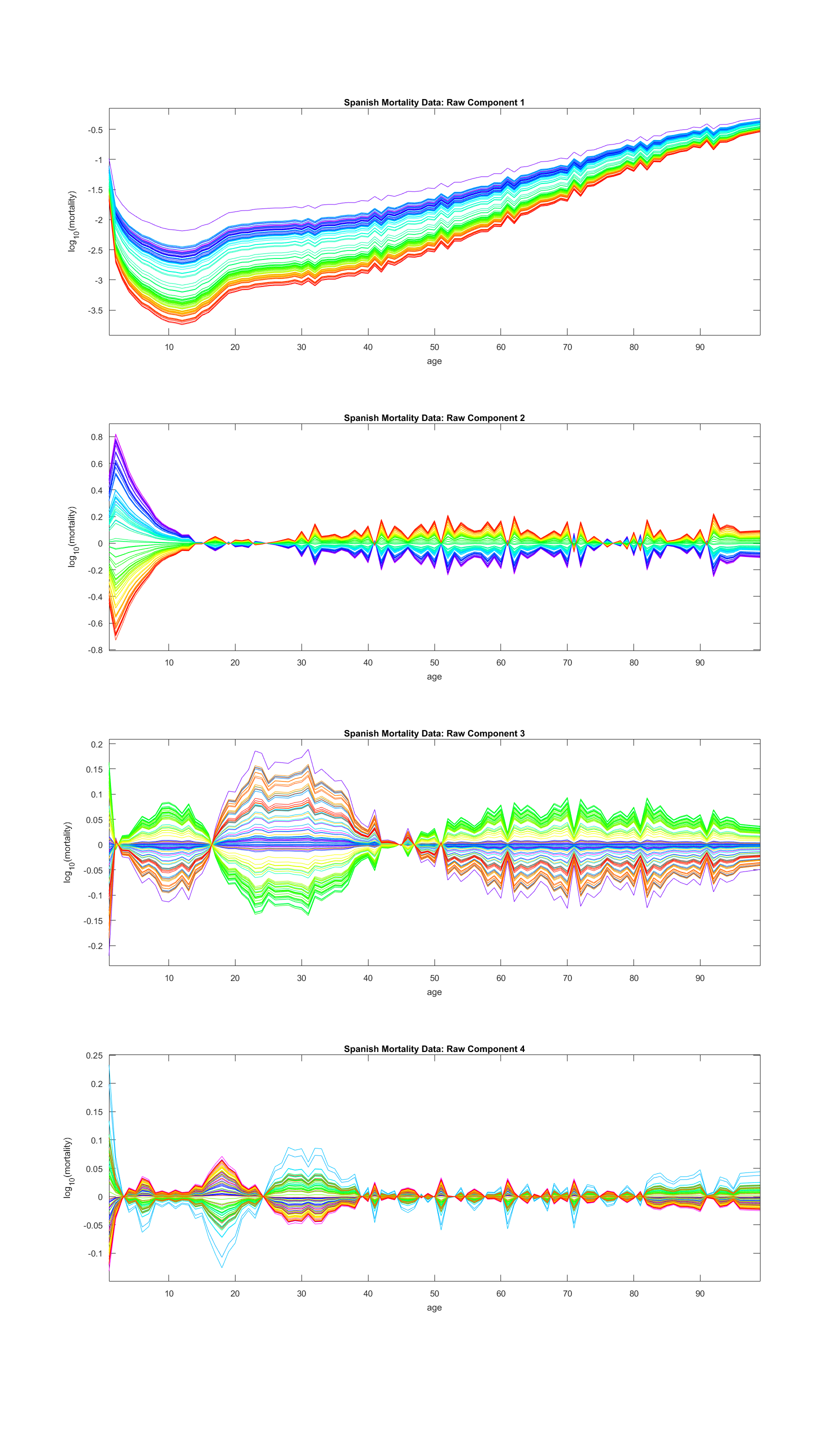}
\caption{Data curve view of uncentered Spanish mortality data. Different centering dramatically changes the visual analytic impression.}\label{fig:mortnoncurves}
\end{figure}

For instance, Figure~\ref{fig:mortnoncurves} shows the modes of variation for the uncentered version of the data matrix. The first component contains information about both the general mortality pattern across ages and the overall improvement over time. The second component has a new contrast between young children and everyone else, and the third component combines many of the patterns separating young adults from older adults with an additional infant effect. Finally, the fourth component reveals a new contrast between younger middle-aged men (ages 25-40) and the rest of the population. The first component contains much of the information taken out by the object mean in the conventional FDA in Figure~\ref{fig:mortobjectcurves}, but it also contains much of what is found in that analysis's first mode of variation.

\begin{figure}[H]
\centering
\includegraphics[scale=0.32]{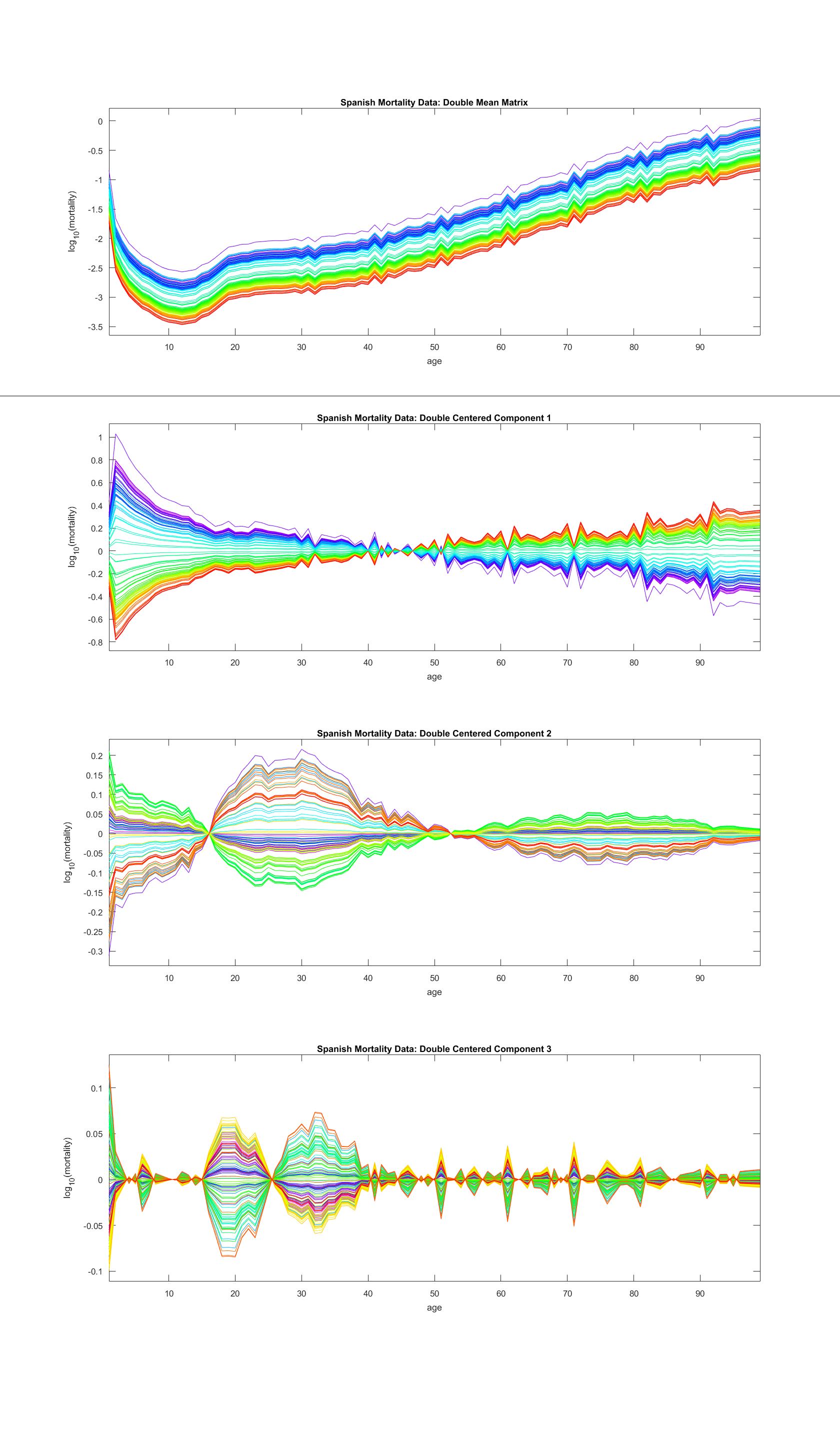}
\caption{Data curve view of double-centered Spanish mortality data. Both the overall improvement over time and differences across ages are contained within the mean, leaving more specific effects for each subsequent mode of variation. The double mean matrix is the sum of the object and trait mean matrices, and is typically rank~2.}\label{fig:mortdoublecurves}
\end{figure}

Next, the double-centered FDA is studied in Figure~\ref{fig:mortdoublecurves}. The rank~2 double mean (first panel) contains both the differences across ages found in the object mean and the constant component of the overall improvement over time found in the trait mean. This visualization of the double mean matrix provides further meaning and context to the first component of the uncentered FDA in Figure~\ref{fig:mortnoncurves}. We can now see that component is a slightly perturbed and lower-rank version of the double mean matrix. Subsequent panels of Figure~\ref{fig:mortdoublecurves} then each show one additional effect, and each panel's effect roughly corresponds to the respective panel from Figure~\ref{fig:mortnoncurves}. The second panel shows stronger improvement over time for younger people, which was also shown in the second panel of the previous figure, but for a more lopsided age group. The third shows differences between the 18-49 year-olds and the rest of the population, which again lines up well with the effect shown in the third panel of the previous figure. The fourth panel shows a difference between older and younger individuals within the 18-49 age range, representing a clearer picture of the contrast hinted at in the fourth panel of Figure~\ref{fig:mortnoncurves}. Each component is cleanly interpretable and untethered from interference due to mean effects. The one aspect of the data still spread throughout components is the age-rounding effect for older individuals, though this happens regardless of the centering chosen.

The distribution of interpretable effects varies uniquely with each form of centering. To summarize the differences, Table~\ref{tbl:mort} displays for each  centering (columns), which phenomena (rows) are contained in which component (numbers). For instance, the first component of the object-centered analysis contains information about both overall mortality improvement and the stronger improvement in mortality rate for younger people, whereas those two phenomena are split up in the double-centered analysis. The former is contained in the mean and the latter is contained in the first component.

\begin{table}[H]
\centering
\begin{tabular}{cccc} \hline
$\downarrow$ Phenomenon, Centering Type $\rightarrow$ & None & Object & Double \\ \hline
Mortality rate differences across ages & 1 & Mean & Object Mean \\ 
Overall mortality improvement & 1 & 1 & Trait Mean\\ 
Stronger improvement for younger people & 2 & 1 & 1\\ 
Contrast between 18-49 and others & 3 & 2 & 2 \\ 
Contrast between 18-25 and 25-40 & 4 &  & 3\\ 
Infant Effects & 3,4 & 3 & 3 \\ 
Age Rounding & 1,2,3 & Mean,1,3 & Object Mean,1,3 \\ \hline
\end{tabular}
\caption{Phenomena in FDA components after different forms of centering. Missing phenomena are indicated by empty cells.} \label{tbl:mort}
\end{table}

The table shows that choice of centering determines in which component different phenomena appear. Different analysts may well have different preferences. We prefer double centering for this data set because it provides the cleanest separation of phenomena into individual modes of variation. Object centering fails to find the additional contrast among the younger adults found with no centering and double centering. While most of the effects are present in the uncentered FDA, double centering allows for clearer attribution of each phenomenon to a specific effect, centering or otherwise. The contrast among younger men is also more prominent and more straightforward in the third mode of variation of the double-centered FDA as compared to the fourth mode in the uncentered FDA.

Often the two most meaningful decompositions into modes of variation will derive from object-centered and double-centered data. In Section~\ref{sec:test}, we present a statistical test to help determine whether object centering or double centering may be more appropriate for a given data set. As will be seen in Section~\ref{sec:formalism}, these two centerings result in mutually uncorrelated score vectors, and double centering additionally results in mutually uncorrelated loadings vectors.

\subsection{Lung Cancer Data}

\begin{figure}[H]
\centering
\includegraphics[scale=0.7]{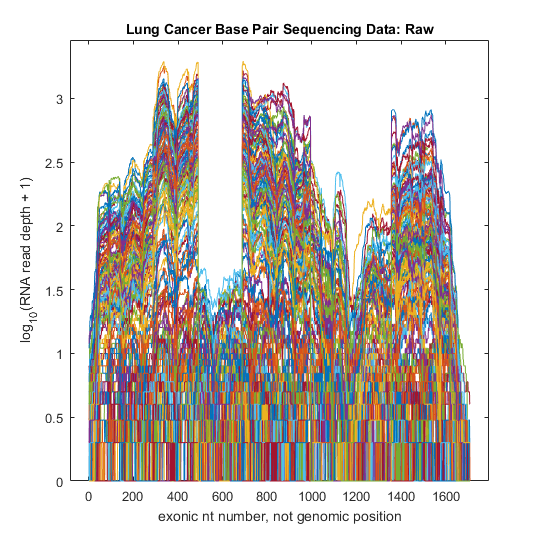}
\caption{Data curve view of lung cancer RNAseq data. Important relationships in the data are hard to discern. The large steps at the bottom are an artifact of the shifted log transformation.}\label{fig:rawLung}
\end{figure}

The default form of centering (usually object centering) can sometimes be the best choice depending on the goals of the analysis. One such situation is clustering  in the context of RNAseq lung cancer gene expression data from \cite{kimes}. Here our data matrix contains 180 observations from lung cancer patients of 1709 base pairs along the gene CDKN2A. Figure~\ref{fig:rawLung} displays the data as a curve bundle. The horizontal axis represents base pair location on the chromosome and for each location the vertical axis displays the $\log_{10}$ of the counts of RNA reads plus 1. Because these reads overlap, traits near one another appear to be strongly correlated.

\begin{figure}[H]
\centering
\includegraphics[width=\textwidth]{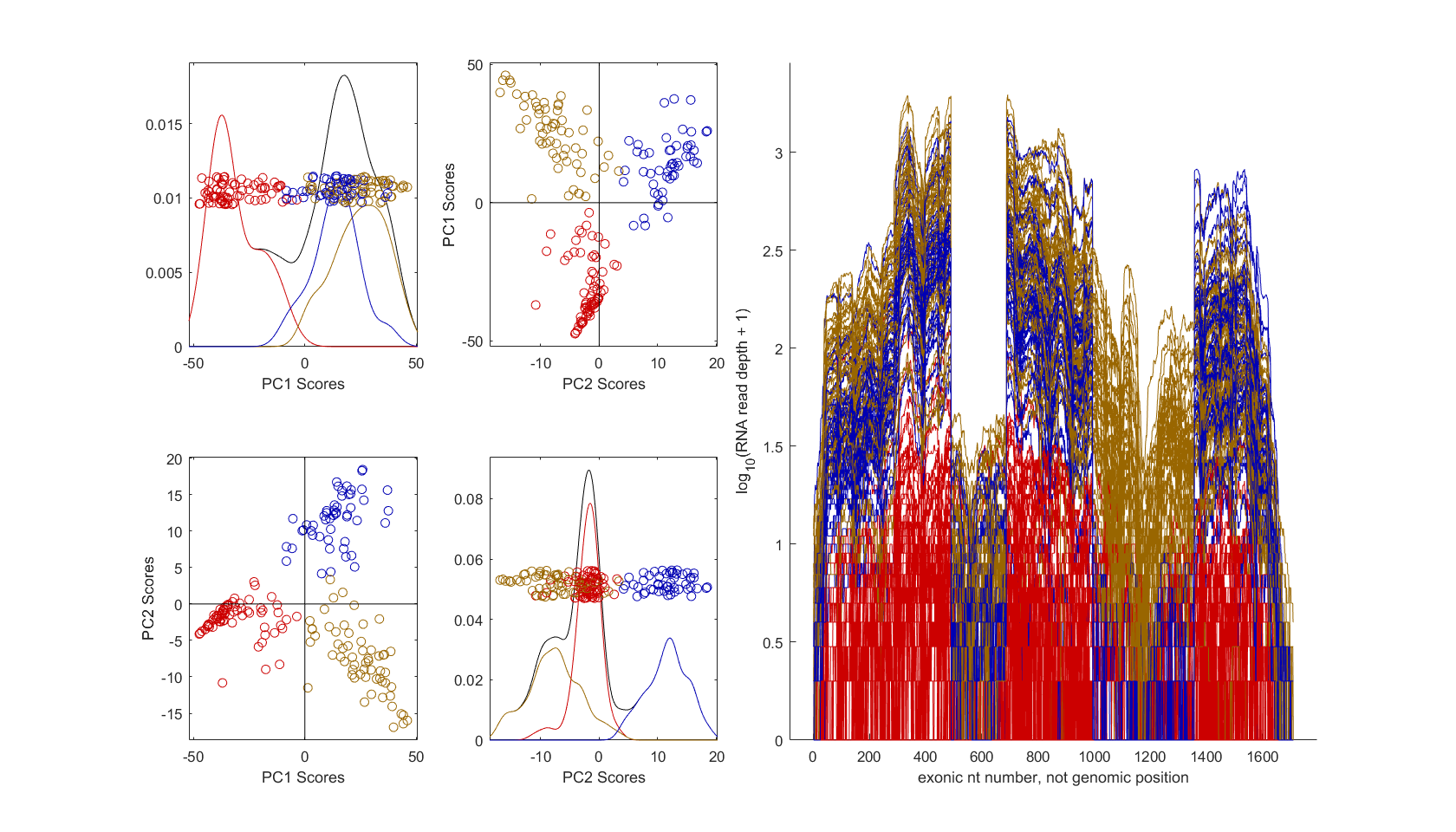}
\caption{(Left) Brushed scores view of traditional FDA of lung cancer base pair RNA expression data. We have three prominent clusters among the first two modes of variation. (Right) Curve view colored with clusters. Red cluster has low expression everywhere, blue \& gold clusters differ between base pairs 1000 and 1400, suggesting alternate splicing as discussed in \cite{kimes}.}\label{fig:objLung}
\end{figure}

To look for relationships within the data, we perform a traditional FDA. In particular, we project the data onto the subspace defined by the first few modes of variation as shown in the left half of Figure~\ref{fig:objLung}. The four panels on the left are 1D and 2D scores plots laid out in a similar format to Figure~\ref{fig:mortobjectscatter}. The first two modes of variation suggest three distinct clusters. We study those clusters via \textit{brushing}: manually coloring data based on visual information in the left side of Figure~\ref{fig:objLung} and then transferring those colors to the curve bundle plot in the right panel of Figure~\ref{fig:objLung}. 

The brushed clusters have a clear, obvious visual interpretation in the curve view of the data. The red individuals have low expression levels across the entire gene (these are classically called \textit{unexpressed}), while the blue \& gold individuals are similar but differ in an important way within the range of base pairs between 1000 and 1400. This event is called \textit{alternate splicing} and is very important in cancer research. Focusing on such differences has led to new discoveries by \cite{kimes}. This data has a clear correspondence between clusters and modes of variation. In particular, the first mode separates the red observations from the others, while the second mode separates blue from gold with red in the middle.

\begin{figure}[H]
\centering
\includegraphics[width=\textwidth]{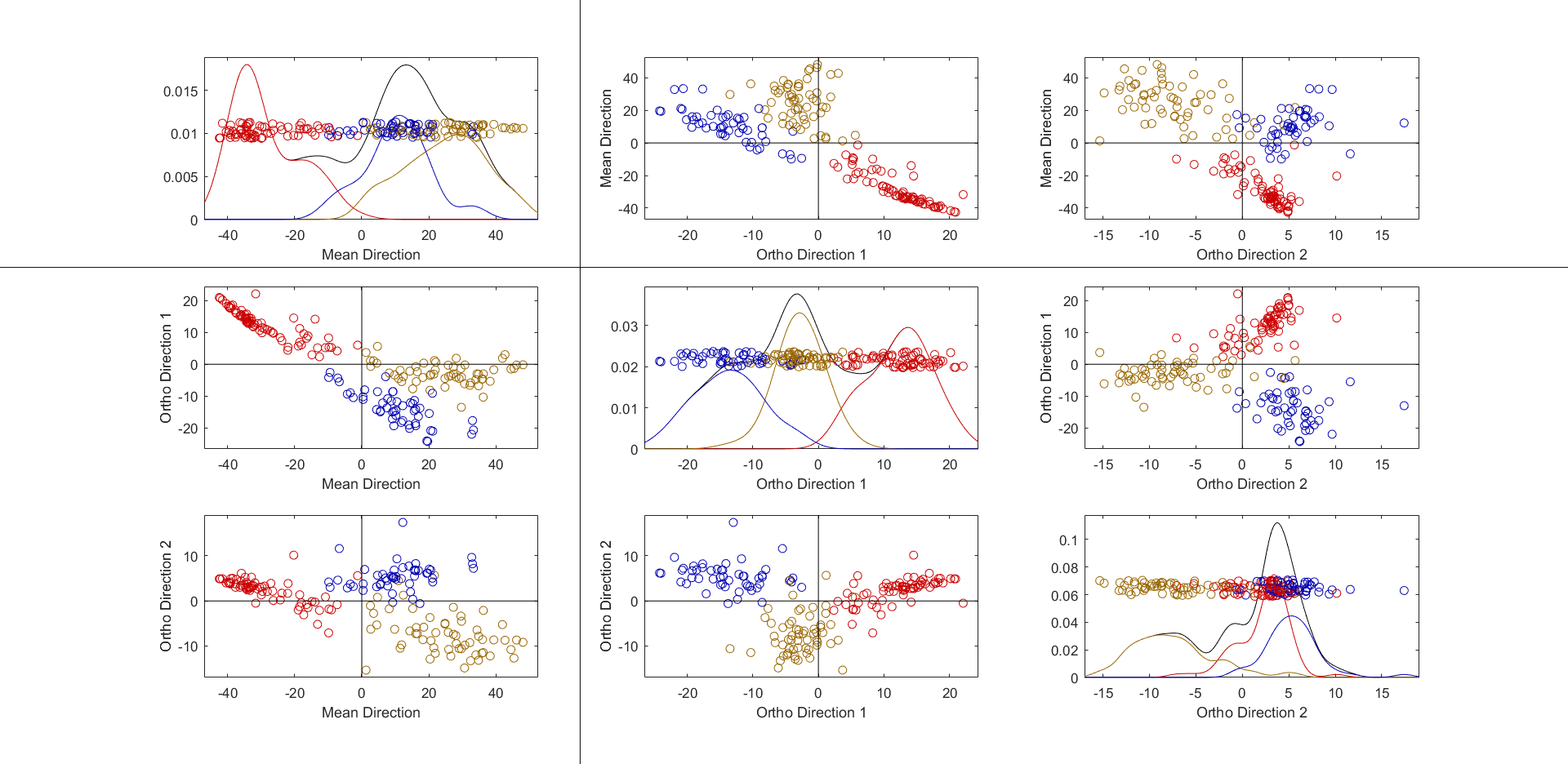}
\caption{(Left) Scores view of double-centered FDA of lung cancer base pair sequencing data. Clusters are made less distinct by involving the trait mean in the visualization.}\label{fig:doubleLung}
\end{figure}

Given this straightforward and interpretable analysis from traditional FDA, what happens when we double-center the matrix instead? Figure~\ref{fig:doubleLung} displays a matrix of 1D and 2D scores plots for the trait mean component and first two orthogonal modes of variation. Note that the three clusters are less visually distinct in these views and the correspondence between modes and clusters is less clear. The separation of the red observations is spread over the trait mean and first orthogonal component, and the separation between blue \& gold is spread across all three directions. Choosing new clusters by brushing this figure would also be much more challenging as no single two-dimensional view shows three clearly distinguished point clouds like those seen in Figure~\ref{fig:objLung}.

In this case, introducing an additional form of centering reduced the interpretability of the results without adding any additional insights. The three-dimensional subspace from the double-centered FDA does no better of a job delineating clusters than the two-dimensional subspace from the typical FDA, and the separation of each group is spread across multiple modes of variation in the double-centered FDA. Opting for double centering over object centering can either enhance interpretability, as in the mortality data, or obscure it, as in the lung cancer data.

\section{Formalism} \label{sec:formalism}


\subsection{Consequences of Different Forms of Centering}

\begin{figure}[H]
\includegraphics[width=\textwidth]{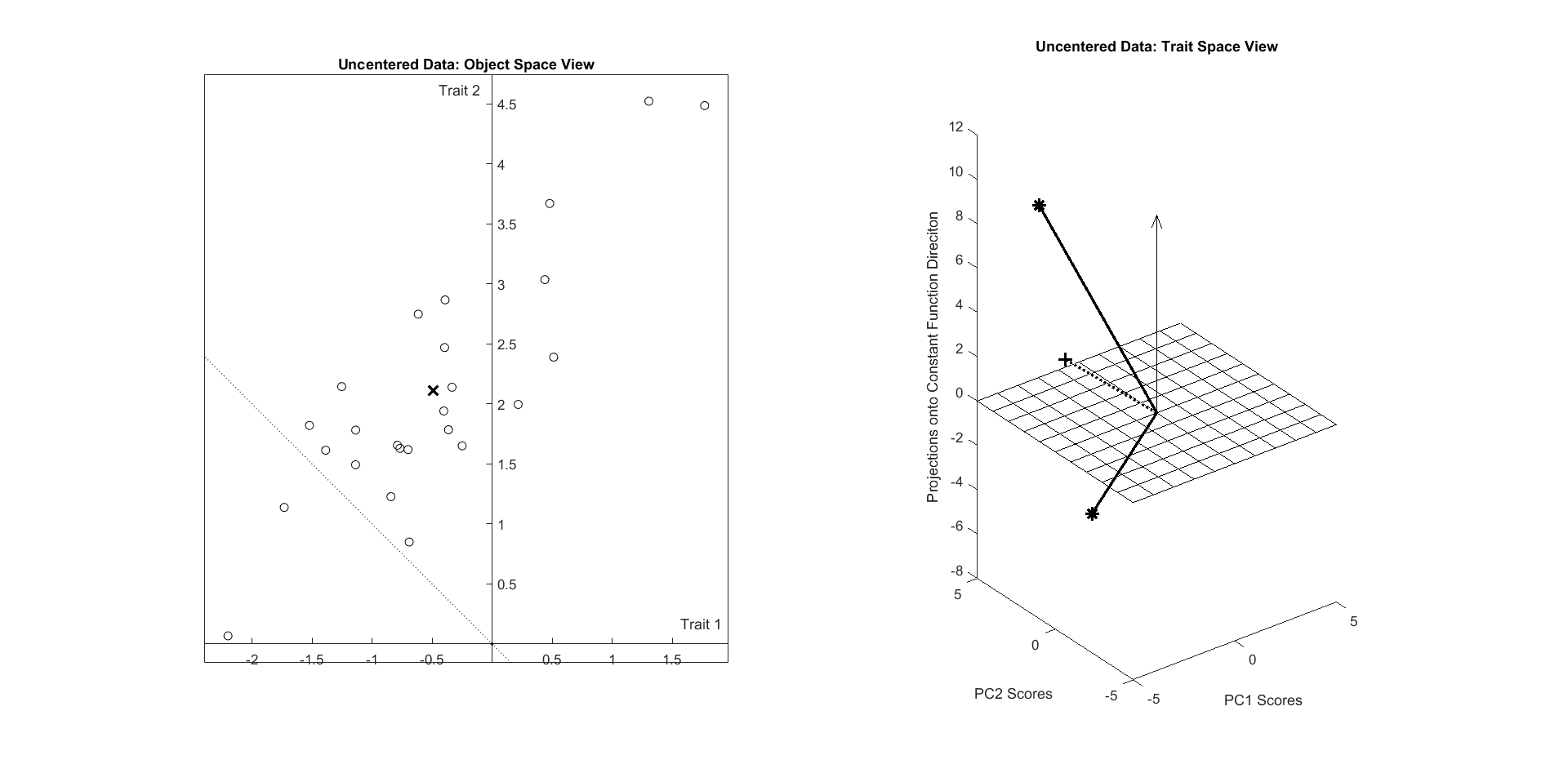}
\caption{Uncentered data matrix $\Xb$ shown in both object space ($\bbR^2$, left) and the three-dimensional subspace of trait space generated by the constant function direction and the data ($\bbR^{25}$, right). Notably, the asterisks and plus sign in the right panel do not lie in the mesh plane orthogonal to the constant function direction (vertical axis).}\label{fig:centnocent}
\end{figure}

We investigate the effects of grand mean, object, trait, and double centering on a small example data matrix $\Xb$ with 2 traits (rows) and 25 data objects (columns). We can then think of $\bbR^2$ as the \textit{object space}, and $\bbR^{25}$ as the \textit{trait space}. Figure~\ref{fig:centnocent} shows the entries of $\Xb$ as they exist in both object space and trait space. The left panel shows the 25 ordered pairs (circles) as a scatter plot in $\bbR^2$. Visualization in $\bbR^{25}$ is more challenging. For studying centering, the \textit{constant function direction}, i.e. vector of 1's, and the subspace it generates are pivotal. Therefore, the right panel shows the two 25-dimensional trait vectors (asterisks) projected into the three-dimensional subspace of $\bbR^{25}$ generated by the constant function direction ($z$-coordinate), and the two orthogonal trait space principal components ($x$ and $y$ coordinates).  Note that the subspace orthogonal to the constant function direction contains every vector whose entries have mean 0. The mesh plane represents the projection of that subspace of $\bbR^{25}$ into the chosen three-dimensional subspace. Also note that in both spaces, the mean vectors of the points ($\times$ in object space, $+$ in trait space) are nonzero. In the right panel, the two data points and their mean are shown as vectors from the origin. In addition to a different symbol, the mean vector is distinguished with a dashed line type.

The following subsections each discuss the results of a different centering on this data. Each subsection has an accompanying figure that visually demonstrates the impacts of each type of centering on the example data matrix in both spaces. Each accompanying figure is formatted similarly to Figure~\ref{fig:centnocent}: the left panel will show object space, and the right panel will show a projection onto the same subspace of trait space. In each subsection we will also discuss how each centering can be interpreted in a third space:  $\bbR^{d\times n}$, the space of $d\times n$ matrices endowed with the Frobenius norm.

\subsubsection{Grand Mean Centering}

We begin our geometric exploration of centering with grand mean centering: the form of centering that finds the scalar grand mean value of all the entries of the matrix and subtracts that value from each entry. 

\begin{figure}[H]
\includegraphics[width=\textwidth]{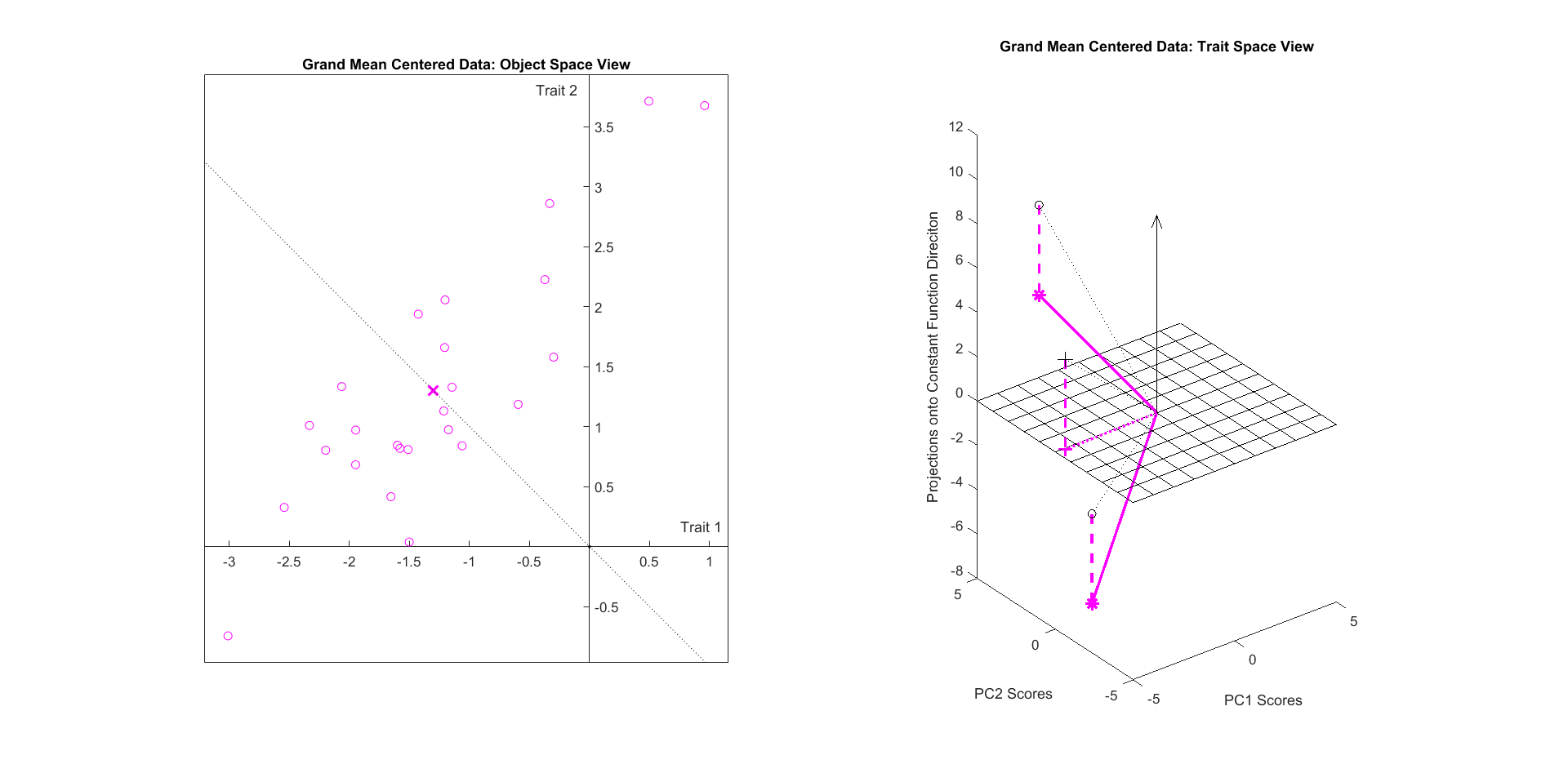}
\caption{Data matrix $\Xb_G$, centered version of $\Xb$ such that all the entries have mean 0, shown in both object space ($\bbR^2$, left) and trait space ($\bbR^{25}$, right). Shows grand mean centering is a translation of point clouds in both spaces.}\label{fig:centgmcent}
\end{figure}

We calculate the \textit{grand mean matrix} $\Mb_G = \1_{d\times n} \mu_G$, where $\mu_G$ is the average of all entries of $\Xb$. The grand-mean-centered version of $\Xb$ is then denoted $\Xb_G = \Xb - \Mb_G$. While this centering is not often performed on its own in data analysis, it serves as an appropriate first step for analyzing the geometric implications of each subsequent centering. Figure~\ref{fig:centgmcent} shows the results of this centering in both object space and trait space, where the point clouds retain their shapes but have been translated to different locations. In both spaces, the data are translated parallel to their corresponding constant function direction such that each mean ($\times$ in $\bbR^2$ and $+$ in $\bbR^{25}$) lies in the subspace orthogonal to their constant function direction.

While we have described the geometric implications of grand mean centering in both object space and trait space, grand mean centering alone is typically not useful in data analysis. The interpretation and consequences of grand mean centering are better studied in $\bbR^{d\times n}$. In this space $\Mb_G$ lies in the constant function direction. Therefore when we subtract $\Mb_G$ from $\Xb$, the resulting matrix $\Xb_G$ is orthogonal to the constant function in the space of matrices. This property then enforces a further orthogonal relationship between the object mean and trait mean matrices, denoted $\Mb_O$ and $\Mb_T$ respectively. We calculate $\Mb_O = \bmu_d \1_n^T$ and $\Mb_T = \1_d \bmu_n^T$, where $\bmu_d$ is the $d$-dimensional vector whose entries are the mean of each trait of $\Xb$ and $\bmu_n$ is the $n$-dimensional vector whose entries are the mean of each data object of $\Xb$. Each of $\Mb_O$ and $\Mb_T$ are rank 1, and $\Mb_O$ has identical columns while $\Mb_T$ has identical rows. If these mean matrices are calculated with respect to $\Xb_G$ rather than $\Xb$, both of these matrices have entries that sum to 0. In this case each column of $\Mb_O$ sums to 0 and each row of $\Mb_T$ sums to 0. This means that with respect to the Frobenius inner product in $\bbR^{d\times n}$, $\Mb_O$ and $\Mb_T$ are orthogonal after grand mean centering.

\subsubsection{Object Centering}

\begin{figure}[H]
\includegraphics[width=\textwidth]{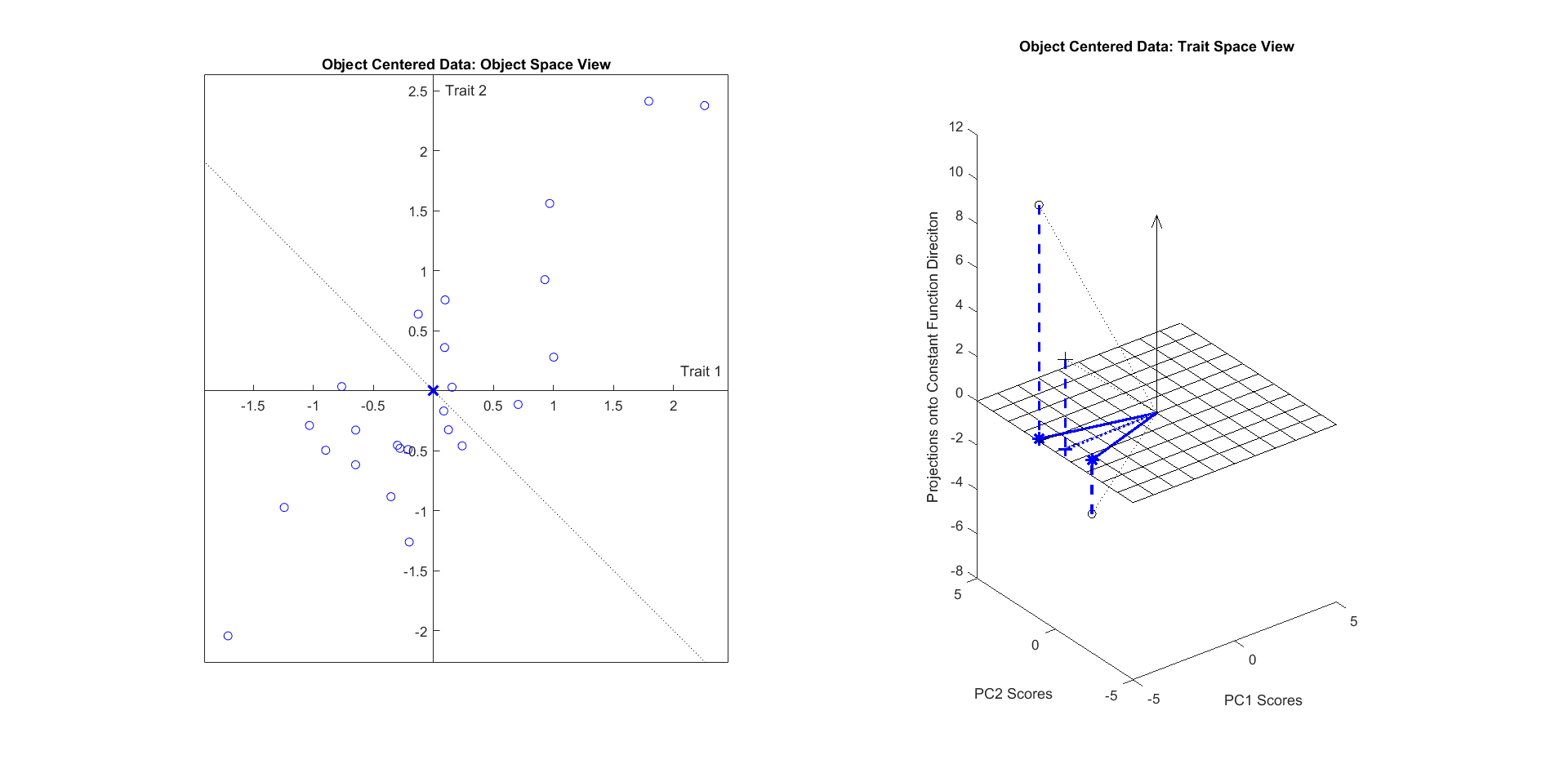}
\caption{Data matrix $\Xb_O$, centered version of $\Xb$ such that the data objects have mean vector $\0$, shown in both object space ($\bbR^2$, left) and trait space ($\bbR^{25}$, right). The trait vectors are projected onto the subspace orthogonal to the constant function direction.}\label{fig:centr2cent}
\end{figure}

Now we explore the centering which is performed on data matrices as a part of typical FDA. We calculate $\Xb_O$, the object-centered version of $\Xb$ such that the data objects have a mean of the $d$-dimensional $\0$ vector. Figure~\ref{fig:centr2cent} shows the results of this form of centering on $\Xb$. As shown in the left panel, the points in object space now have a mean vector at exactly the origin. The points have all been translated from their locations in Figure~\ref{fig:centnocent} by the same amount and in the same direction. The two trait vectors undergo a different transformation in $\bbR^{25}$. Each vector, as well as their vector mean, is projected into the 24-dimensional subspace which is orthogonal to the constant function direction. This 24-dimensional subspace is again represented by the mesh plane.

To facilitate a PCA-like decomposition into modes of variation of $\Xb_O$, consider an SVD of $\Xb_O=\Ub_O\Db_O\Vb_O^T$. In our matrix orientation convention, $\Ub_O$ is associated with loadings and $\Vb_O$ is associated with scores. As the trait vectors are rows of $\Xb_O$ and therefore now lie in the subspace orthogonal to the constant function direction, the orthonormal basis for their span, i.e. the columns of $\Vb_O$, will also be composed of vectors orthogonal to the constant function direction. These entries represent the scores of each observation along each direction, and centering this way guarantees that each set of scores has mean 0. 

\subsubsection{Trait Centering}

\begin{figure}[H]
\includegraphics[width=\textwidth]{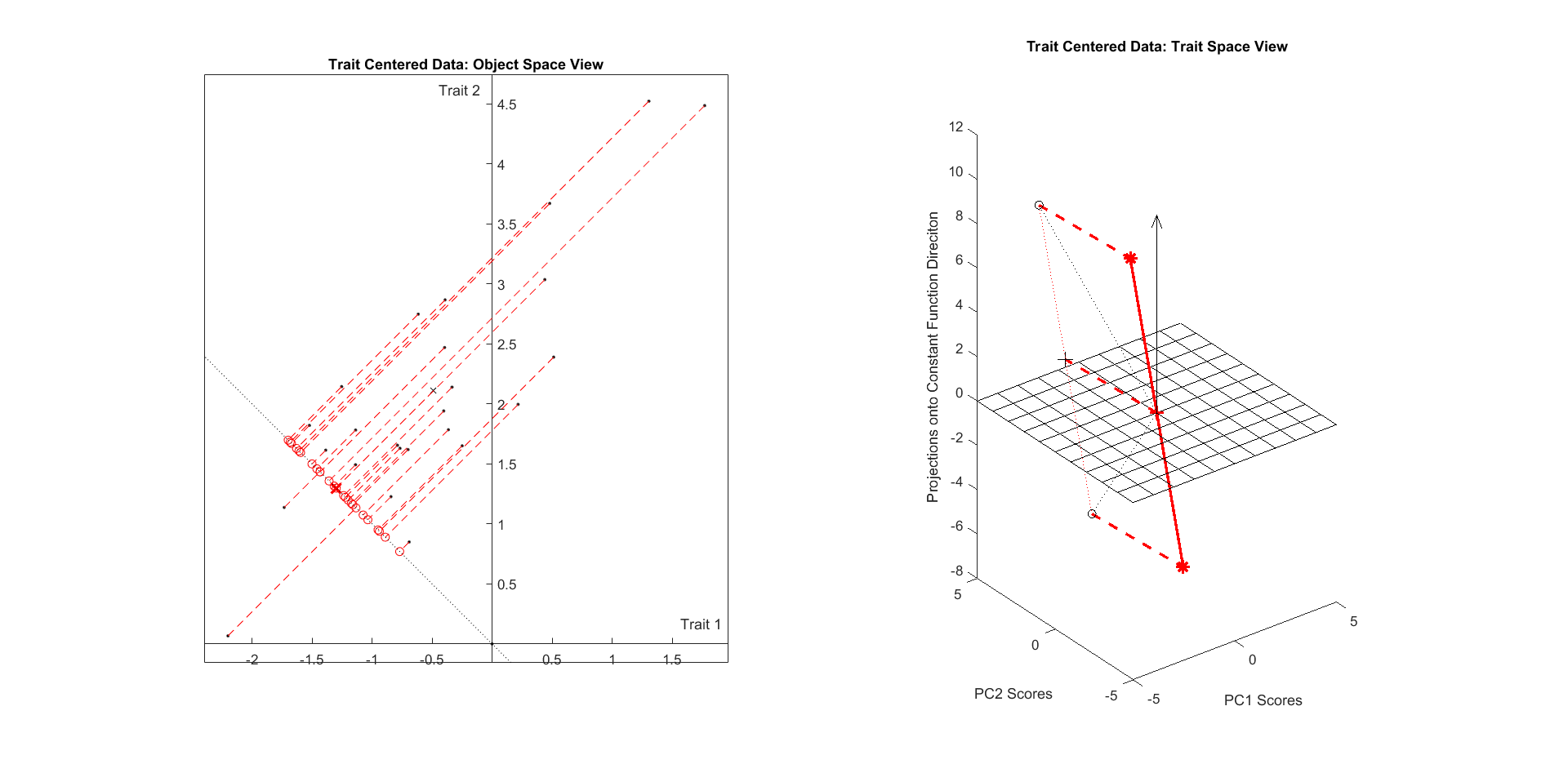}
\caption{Data matrix $\Xb_T$, centered version of $\Xb$ such that the traits have mean vector $\0$, shown in both object space ($\bbR^2$, left) and trait space ($\bbR^{25}$, right). As a consequence, the objects are projected onto the subspace orthogonal to the respective constant function direction. This projection demonstrates that $\Xb_T$ is of lower rank than $\Xb$.}\label{fig:centrncent}
\end{figure}

The second centering is the dual of the centering used in PCA. We calculate $\Xb_T$: the centered version of $\Xb$ such that the traits have a mean of the $n$-dimensional $\0$ vector. Figure~\ref{fig:centrncent} shows the results of this centering on the data matrix $\Xb$ from Figure~\ref{fig:centnocent}. The left panel shows that the points in object space have been projected onto the subspace orthogonal to the $\bb{R}^2$ constant function direction, while the right panel shows that the points in trait space have been translated such that their mean is at exactly the origin. This result is of course the dual of the previous form of centering. Note that the resulting matrix $\Xb_T$ is now rank 1 instead of rank 2. 

To similarly find the modes of variation of $\Xb_T$, consider an SVD of $\Xb_T=\Ub_T\Db_T\Vb_T^T$. As the data object vectors now lie in the subspace orthogonal to the constant function direction in $\bbR^2$, the orthonormal basis for their span (columns of $\Ub_T$) will also be composed of vectors in this subspace. These entries represent the unweighted loadings of each trait within each mode of variation of the data objects, and centering this way guarantees that each set of loadings has mean 0. 

\subsubsection{Double Centering}

\begin{figure}[H]
\includegraphics[width=\textwidth]{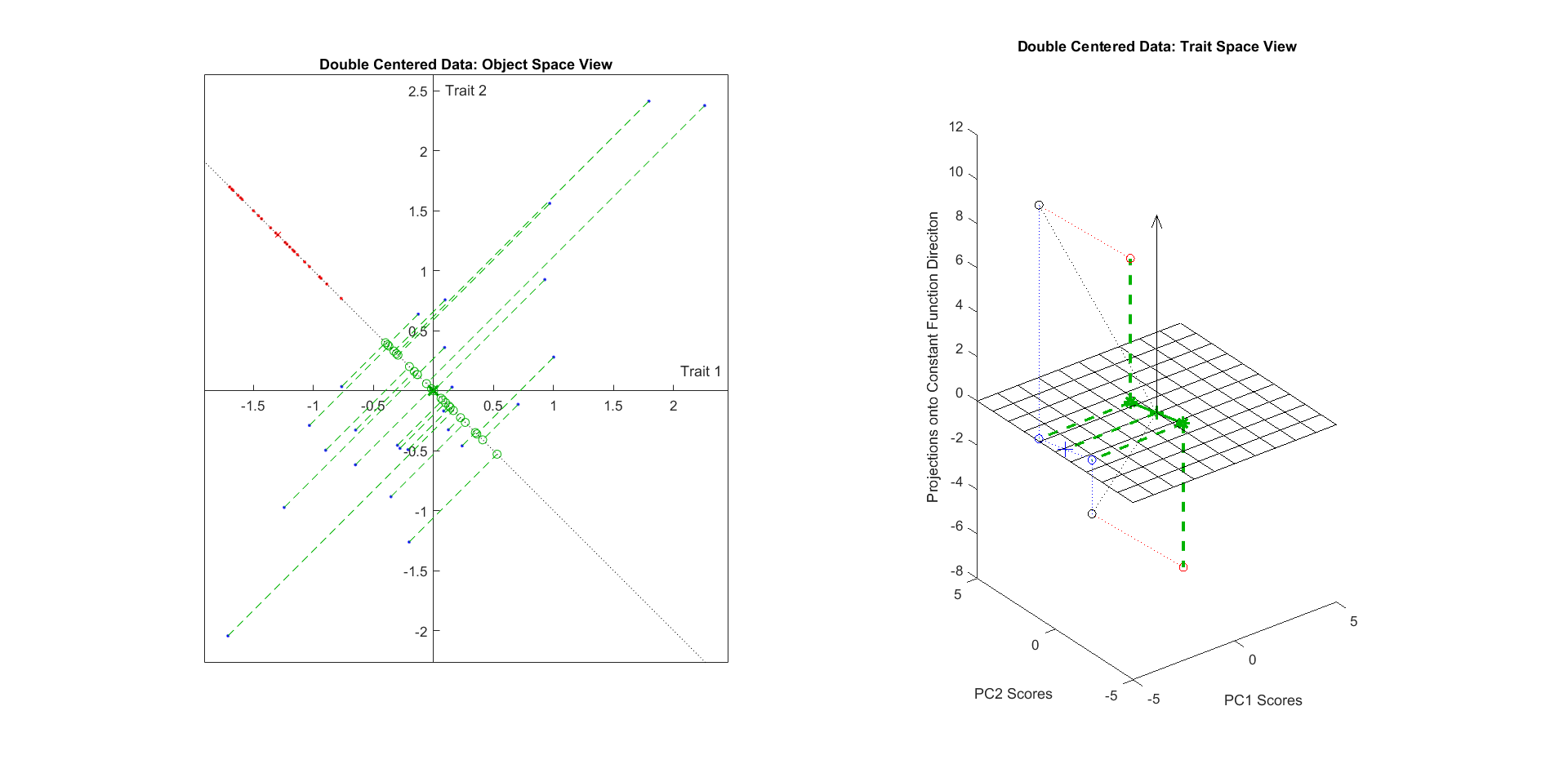}
\caption{Data matrix $\Xb_D$, centered version of $\Xb$ such that the data objects and traits both have mean vector $\0$, shown in both object space ($\bbR^2$, left) and trait space ($\bbR^{25}$, right). As a consequence, both the data objects and traits are projected onto respective subspaces orthogonal to the constant function direction.}\label{fig:centbothcent}
\end{figure}

The final mode of centering combines the operations of both previous forms into a single transformation. We calculate $\Xb_D$, the double-centered version of $\Xb$ where the traits have a mean of the $n$-dimensional $\0$ vector and the data objects have a mean of the $d$-dimensional $\0$ vector. Figure~\ref{fig:centbothcent} shows the results of double centering the matrix $\Xb$ from Figure~\ref{fig:centnocent}. In both panels, the original points have been translated so that their mean lies at the origin and they are projected onto the subspace orthogonal to the corresponding constant function direction. The double-centered ordered pair objects in the left panel and corresponding double-centered trait vectors are all shown in green.

Note that these two operations, projection and translation, are commutative. Projecting the first-translated points results in the same transformed data as translating the first-projected points. We can see this commutation in both panels of Figure~\ref{fig:centbothcent}. In the left panel we can arrive at the green points either by projecting the previously-translated blue points onto the line orthogonal to the constant function direction or by translating the previously-projected red points such that their vector mean now lies at the origin. In the dual situation in the right panel, we can arrive at the green points either by projecting the previously-translated red points onto the mesh plane indicating the subspace orthogonal to the constant function or by translating the previously-projected blue points such that their vector mean now lies at the origin.


This matrix is also of lower rank than $\Xb$.

To similarly find the modes of variation of $\Xb_D$, consider an SVD of $\Xb_D=\Ub_D\Db_D\Vb_D^T$. Both the data objects and trait vectors lie in subspaces orthogonal to their respective constant function direction, so both the sets of loadings and the sets of scores for this data will have mean 0.

\subsection{Discussion}

There is a strong connection between mutual orthogonality of vectors and correlation in corresponding scatterplots that is driven by the means of the entries of each vector.  For two observed unit vectors $\xb$ and $\yb$, because $\sum x_i^2 = \sum y_i^2 = 1$, the correlation of their entries is $Corr(\xb,\yb) = \sum(x_iy_i) - \sum(x_i)\sum(y_i)$. Since FDA scores and loadings vectors are mutually orthogonal in their respective spaces, we will always have $\sum (x_i y_i) = 0$. Therefore a sufficient condition for the entries of two scores and/or loadings vectors to be \textit{uncorrelated} is for the entries of one vector to have mean zero. 

Table~\ref{tbl:uncorr} summarizes how this condition enforces uncorrelatedness in scores and loadings vectors found via FDA of differently-centered matrices. Whether FDA of a single matrix is treated as an eigenanalysis of a covariance matrix or as an SVD of a (possibly centered) data matrix, the scores and loadings vectors will always be mutually orthogonal. This fact combined with the projection operations involved in different centerings can produce mutually uncorrelated scores and/or loadings vectors. This uncorrelatedness is most prominent and important when forming scatter plots like those shown in Figure~\ref{fig:mortobjectscatter}.

\begin{table}[H]
\centering
\begin{tabular}{ccccc} \hline
$\downarrow$ Effect, Centering Type $\rightarrow$ & None & Object & Trait & Double \\ \hline
Orthogonal Score Vectors & $\checkmark$  & $\checkmark$ & $\checkmark$ & $\checkmark$ \\ 
Uncorrelated Score Vectors &  & $\checkmark$ & & $\checkmark$ \\ 
Orthogonal Loadings Vectors & $\checkmark$ & $\checkmark$ & $\checkmark$  & $\checkmark$ \\ 
Uncorrelated Loadings Vectors &  & & $\checkmark$  & $\checkmark$ \\ \hline
\end{tabular}
\caption{Summary of which centerings produce which outcome for sets of scores and loadings vectors in FDA.} \label{tbl:uncorr}
\end{table}

As a remark, some of the centerings resulted in loss of rank in our synthetic data matrix. Recall that the matrix was $2\times 25$, and the matrix became rank~1 after trait centering and double centering. The centerings that involved translation in $\bbR^{25}$, and therefore projection in $\bbR^{2}$, were the ones that reduced the rank of the matrix. In general, the centering that involves projection in the lower-dimensional space out of the trait vector and object vector spaces will result in loss of rank.

\section{Quantifying Double Centering in Functional Data Analysis} \label{sec:test}

As discussed in Sections~\ref{sec:fda} and \ref{sec:formalism}, object centering is the standard default for FDA. This is recommended because interesting structure is often found in variation about that mean vector so its dominating effect is removed and treated separately. As seen in the transition between the left panels of Figures~\ref{fig:centnocent} and \ref{fig:centr2cent}, subtraction of the object mean results in a translation of the data objects in object space $\left(\bbR^d\right)$ so their mean vector becomes the origin. The FDA modes of variation among the now-translated point cloud are then readily calculable via SVD.

However, in cases like the Spanish Mortality data studied in Section~\ref{sec:fda}, an additional dominating effect due to the trait mean can remain within the point cloud even after removal of the object mean. In object space, the trait mean manifests through projection onto the constant function direction. Each entry of the trait vector mean is the signed magnitude of the projection of a corresponding data object vector onto the constant function direction. If a substantial proportion of the object-centered point cloud energy lies along the constant function direction, the trait mean effect may be concealing more interesting structure.

In this section we develop a \textit{direction-energy} hypothesis test to determine when the proportion of energy in the constant function direction becomes "substantial" enough to warrant potential separate consideration of the trait mean mode of variation and the remaining (double centered) modes. In particular, this separation is warranted when the energy proportion along the constant function direction (after object centering) is larger than what is typical for this data. For this, we select $B$ $d$-dimensional unit vectors in object space uniformly at random from the subspace generated by the data, and calculate the proportion of total point cloud energy that lies in each direction. We compare the energy in the constant function direction to the empirical null distribution of energies in random directions. If the constant function energy proportion is a high percentile (95th or above) of that empirical null distribution, we reject the null hypothesis that the energy congregated around the constant function direction is there due to random chance. If there is systematic variation near the constant function, it is often better to remove the trait mean and treat it separately from other modes of variation.

\begin{figure}[H]
\centering
        \begin{minipage}[b]{0.49\linewidth}
            \centering
            \includegraphics[width=\textwidth]{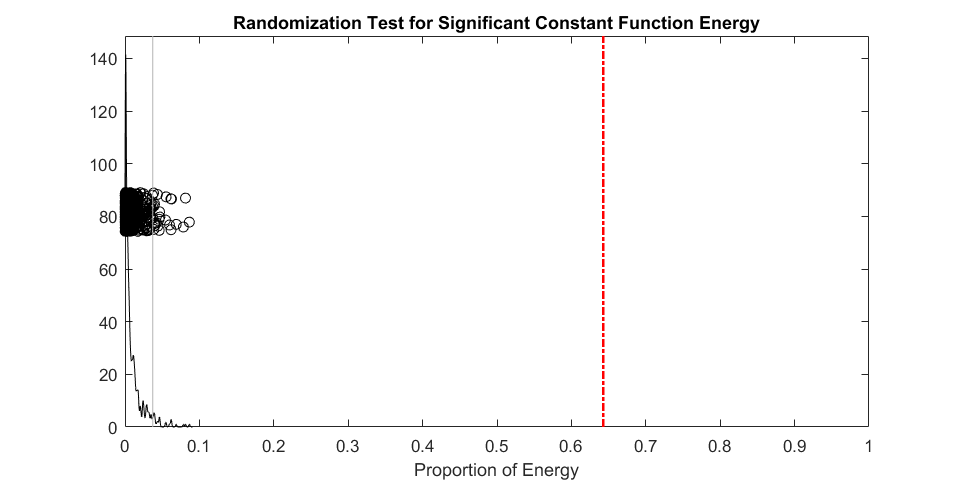}
            
        \end{minipage}
        \begin{minipage}[b]{0.49\linewidth}
            \centering
            \includegraphics[width=\textwidth]{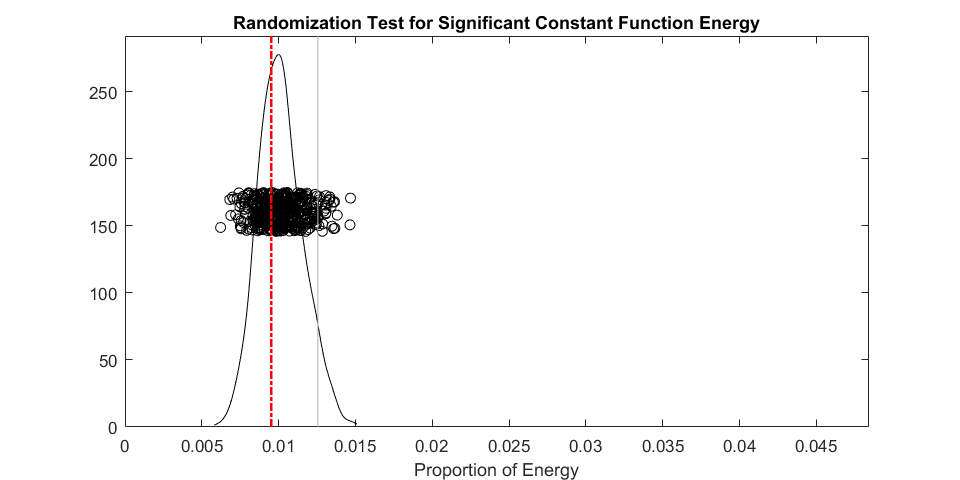}
            
        \end{minipage}
        \caption{Left Panel: Direction Energy Hypothesis Test on Spanish Mortality data. Energy proportion in constant function is much larger than what would be expected due to random chance. Right Panel: Direction energy hypothesis test on $n=100$ synthetic 100-dimensional Gaussian observations. Energy proportion in constant function direction is not distinct from empirical null distribution.} \label{fig:simpleenergy}
\end{figure}

Figure~\ref{fig:simpleenergy} visually displays the results of the hypothesis test described above for two data sets. The left panel shows the test as administered to the mortality data, and the right panel shows the test administered to a synthetically generated matrix of 100 observations from a 100-dimensional standard normal distribution. We plot energy proportion on the horizontal axis with the vertical axis representing density of the randomly generated energy proportions The black dots in each panel represent the $B=500$ energy proportions of each randomly chosen direction. The black curve in each panel is a smooth histogram representing the empirical null distribution of random direction energy proportions. The red dot-dash line in each panel represents the energy proportion in the constant function direction. For the Spanish mortality data in the left panel, nearly 65\% of the energy is congregated around the constant function direction, as is apparent from Figures~\ref{fig:mortobjectcurves}, \ref{fig:mortnoncurves}, and \ref{fig:mortdoublecurves}. This proportion is much higher than that in any random direction, whose energy proportions are shown with the black circles. This result indicates that substantial gains in interpretability are possible for this data set via opting to double-center the data matrix before performing FDA, as demonstrated in Figure~\ref{fig:mortdoublecurves} and Table~\ref{tbl:mort}. Contrastingly, in the right panel, the energy proportion in the constant function direction is not remarkable in any way in the spherically-symmetrical synthetic data displayed.

\begin{figure}[H]
\centering
        \begin{minipage}[b]{0.49\linewidth}
            \centering
            \includegraphics[width=\textwidth]{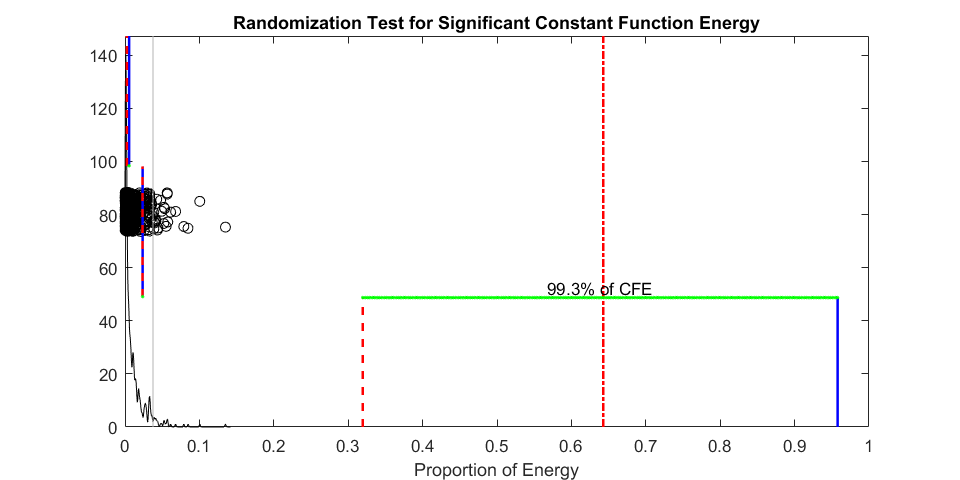}
            
        \end{minipage}
        \begin{minipage}[b]{0.49\linewidth}
            \centering
            \includegraphics[width=\textwidth]{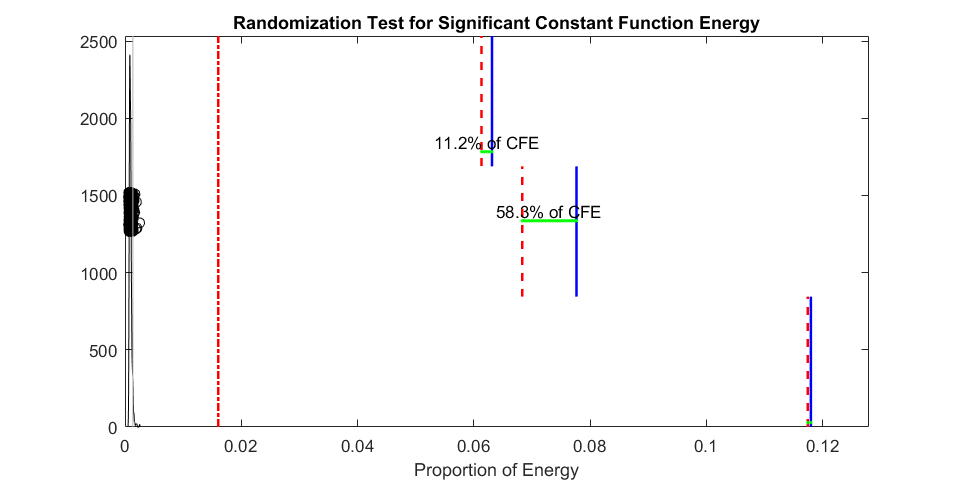}
            
        \end{minipage}
        \caption{Direction energy hypothesis tests on Spanish Mortality (left) and Breast Cancer RNAseq (right) with energy breakdowns by FDA component. Left Panel: Constant function direction almost entirely takes energy from the first component. Right Panel: Constant function direction contains significant information, but its spread across many components and its total energy share is small relative to the first several components.} \label{fig:energybreakdown}
\end{figure}

We can further study the effects of removal of the trait mean by plotting the energies in each FDA component before and after double centering. Figure~\ref{fig:energybreakdown} shows such a breakdown for the Spanish mortality data (left panel) and breast cancer RNAseq data studied in \cite{TCGA/PanCan} (right panel). The solid blue lines show how much energy is accounted for in each object-centered FDA component; components are shown in the order they're found from bottom to top in the figure. The red dashed lines show how much energy is accounted for in each double-centered FDA component; and they're displayed in a similar fashion to the blue lines. All energy proportions are in terms of the total energy in the object-centered data matrix, so the constant function direction energy is included in the total for the double-centered FDA. Consequentially, there is less energy to be allocated for the double-centered components. The blue lines in the left panel correspond with the components shown in Figure~\ref{fig:mortobjectcurves} while the red dashed lines in the left panel correspond with the components shown in Figure~\ref{fig:mortdoublecurves}. 

In the object-centered FDA of the mortality data, the first component accounted for more than 95\% of the energy in the data, but after double centering that energy is split between the constant function direction and the first orthogonal component. In fact the drop in energy share of the first component between object-centered \& double-centered FDA accounts for 99.3\% of the energy share of the constant function direction. This corresponds with the interpretation of this operation in Section~\ref{sec:fda}, where the first object-centered FDA component contained information about both overall improvement and greater improvement for young people, while the first double-centered FDA component is only about greater improvement for young people as the overall improvement is sequestered to the constant function mode. 

The effect of the constant function direction is much less pronounced in the RNAseq data. While it appears to include a statistically significant amount of the overall matrix energy, its energy proportion is still trumped by those of several orthogonal principal components, including the three shown in the right panel. This is likely because a procedure of a similar flavor to removal of the trait mean has already been performed on this data. The columns of this data are normalized such that each has an identical upper quartile. While this operation doesn't entirely remove the effect of the trait mean, it still removes much of the variation in the data objects not explained by the traits.

\section{Centering in Data Integration} \label{sec:pls}

In data integration tasks, two or more data matrices that share common row and/or column dimensions are analyzed in tandem. The goals of such tasks are to reveal what information is shared between the data matrices and to combine the information from each data block to arrive at a more complete picture of the population and/or traits in question. Data matrix centering can play an even more complex role in these situations than in single-matrix FDA as both data objects and traits play critical roles in the analysis. Depending on the chosen methodology, different centering choices may affect the outputs in surprising ways.

We explore the integrative analysis of two data blocks using partial least squares (PLS) under different centering regimes. We choose PLS as a simple and direct method that takes into account potential scaling differences between data blocks. 

\subsection{Partial Least Squares} \label{subsec:pls}

PLS as a data integration procedure derives shared information from the \textit{cross-covariance matrix} $\Sigmab_{1,2}$ between the two data blocks. Here, the cross-covariance matrix refers to the submatrix of the grand covariance matrix of all traits in either data block associated with the covariances between traits across blocks:
$$ \textbf{cov} \left(\left[\begin{matrix}
\Xb_1 \\ \Xb_2
\end{matrix}\right]\right) = \left[\begin{matrix}
\Sigmab_1 & \Sigmab_{1,2} \\ \Sigmab_{1,2}^T & \Sigmab_2
\end{matrix}\right]$$
We can also find the cross-covariance matrix by multiplying object-centered versions of the two data blocks together: $\Sigmab_{1,2} = \frac{1}{n}\Xb_{1O} \Xb_{2O}^T$. As PLS operates on covariance matrices, it chooses pairs of score vectors for each data block with maximal covariance between them.

Importantly, as discussed in \cite{plsoverview}, different variations of PLS lead to different centering-based consequences. As is the case in many data integration methods, one piece of information, either scores or loadings, is calculated first while the other is found subsequently with a projection operation involving the first piece of information and the original data blocks. Whichever set of vectors is found first will be predictably affected by centering choices during preprocessing of the data blocks, but the subsequently found set of vectors are typically not even mutually orthogonal due to the projection.

One approach is to directly take a singular value decomposition of the cross-covariance matrix; the resulting left and right singular vectors then constitute the estimated loadings vectors for $\Xb_1$ and $\Xb_2$ respectively. This results in loadings vectors that are uncorrelated only when the data blocks are double centered. As per Table~\ref{tbl:uncorr}, trait centering and double centering are the two choices that result in loadings vectors with uncorrelated entries. We do not consider trait centering as a possible choice since object centering is required to correctly form the cross-covariance matrix in the first place.

Another approach is to sequentially and algorithmically calculate each score vector, then its corresponding loadings vector, then remove the one-dimensional subspace approximation defined by those vectors before searching for subsequent scores and loadings vectors. As this procedure calculates score vectors first, we can guarantee that the calculated score vectors will be uncorrelated due to object-centering the data blocks. We opt for this approach to mirror other data integration methods that first locate score vectors, including canonical correlations analysis (CCA) from \cite{cca} and angle-based joint and individual variation explained (AJIVE) from \cite{ajive}.

\subsection{Synthetic Data Example}

\begin{figure}[h]
\includegraphics[width=\textwidth]{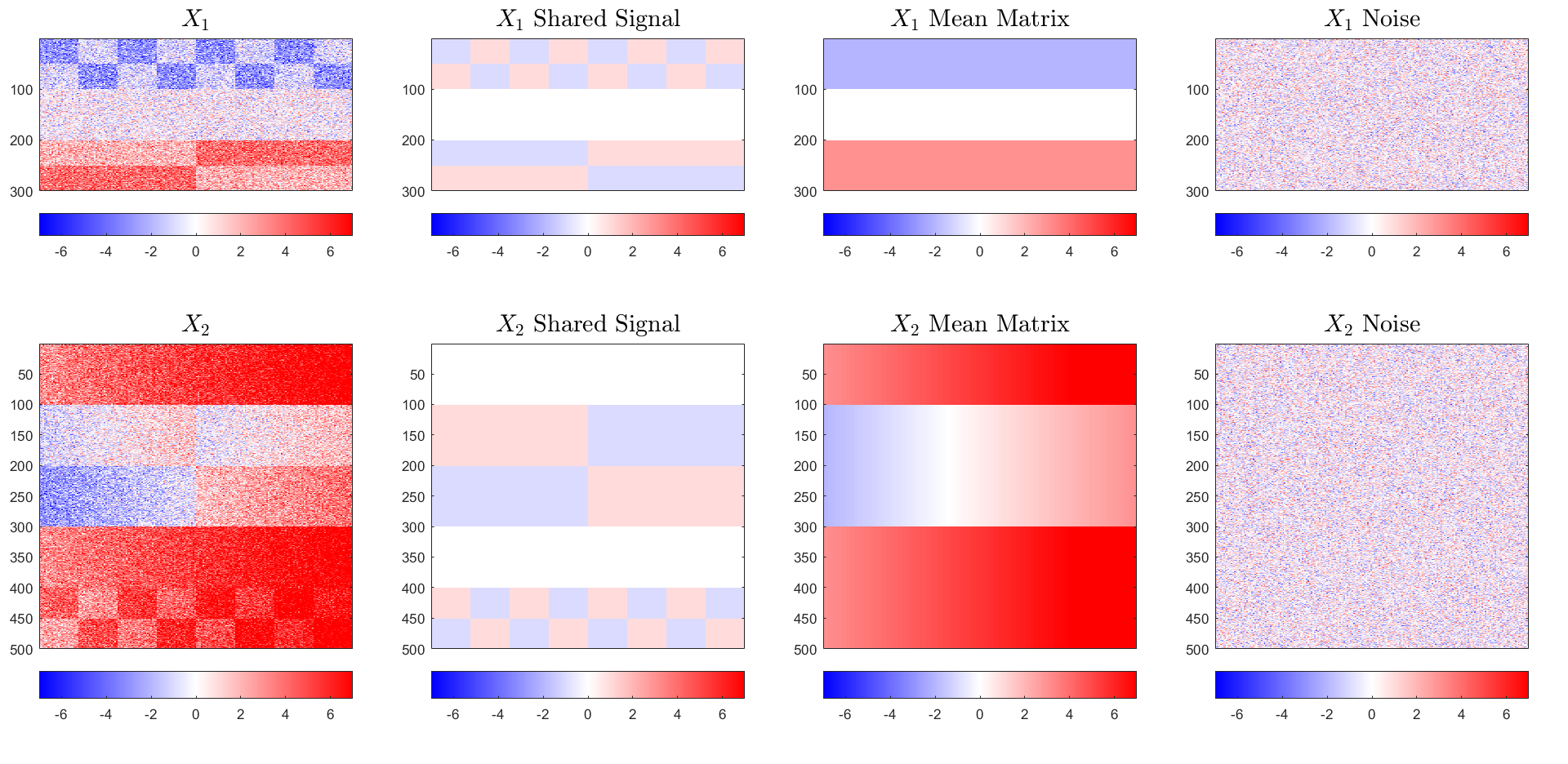}
\caption{Three stages of synthetic data example construction. Underlying rank-two signal (left), underlying signal with added mean effect perturbation (middle), noise perturbation (right).}\label{fig:plstoy}
\end{figure}

To demonstrate the additional complexities involved in centering choice for data integration, we first use the synthetic two-block data set shown in Figure~\ref{fig:plstoy}. The first block, $\Xb_1$, is $300\times 200$, and the second block $\Xb_2$, is $500\times 200$. Note that each block has the same number of data objects (columns) but different numbers of traits (rows). For example, one could represent demographic data and the other could represent various biomarker observations about a cohort of patients. Each data block is formed by adding a rank-two signal matrix to a full-rank Gaussian noise matrix. The underlying components of each signal matrix lie in the same common subspace of trait space, representing shared information between the blocks. However, the overlapping subspaces are obscured by object and trait mean effects in each matrix.

Figure~\ref{fig:plstoy} uses heatmaps to display the construction of the synthetic data example we will use to demonstrate the value of exploring double centering in data integration contexts. The left panels show the observed data matrices, which are formed by additively combining the other matrices in each respective row. The heatmaps in the second column display a shared, underlying rank-two signal in both $\Xb_1$ and $\Xb_2$. By construction, this underlying rank-two joint signal is double centered. The heatmaps in the third column show the mean effects added to each matrix. The matrix added to $\Xb_1$ is rank 1 and represents an object mean matrix as each column is identical. The matrix added to $\Xb_2$ is rank 2 and represents a double mean matrix with both object mean and trait mean components. Finally, the heatmaps in the fourth column display the i.i.d. Gaussian noise added to the observations. The color scale is kept constant across all heatmaps in Figure~\ref{fig:plstoy} to appropriately convey differences in effect size between the shared signal and mean effects.

The object mean vector added to columns of $\Xb_1$ increased the values in the top 100 rows and decreased the values in the bottom 100 rows. An object mean vector was added to $\Xb_2$, but its visual impression is swamped by that of the trait mean effect. The trait mean vector has entries that gradually increase from the first observation's entry to the last. This creates the color gradient visual effect seen in the third panel of the second row.

\begin{figure}[h]
\includegraphics[width=\textwidth]{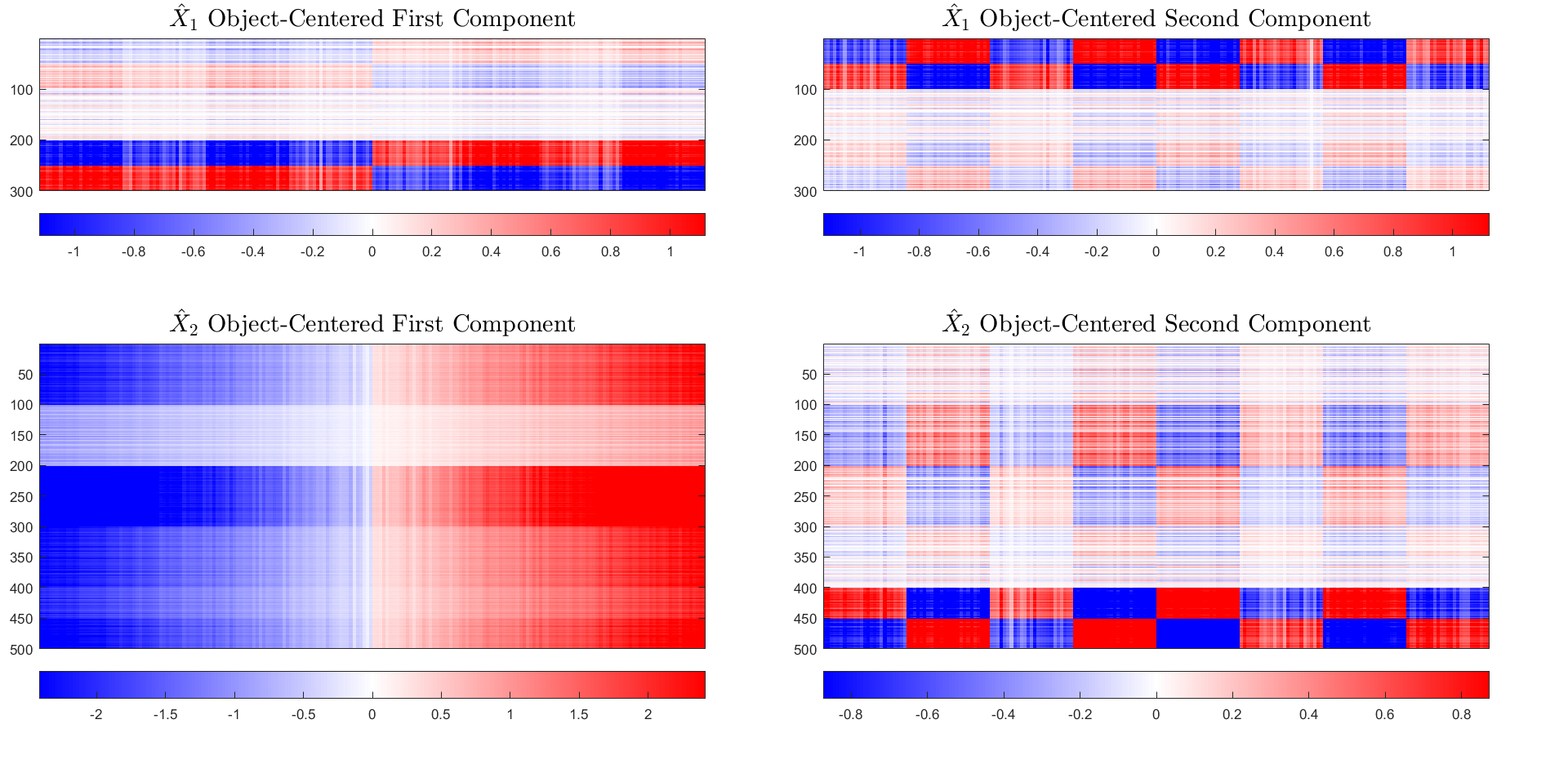}
\caption{First two PLS components of each object centered synthetic data block. Recovery and parsing of distinct underlying signal pieces is reasonable for $\Xb_1$ but the trait mean effect dominates the first component of $\Xb_2$.}\label{fig:plsobj}
\end{figure}

We perform PLS on this two block data set after object centering and after double centering. Figure~\ref{fig:plsobj} displays the first two PLS components of each data block found using the object-centered versions of the matrices. The top panels show the $\Xb_1$ components and the bottom panels show the $\Xb_2$ components. Each estimated $\Xb_1$ component roughly corresponds with one of the rank one underlying signal components shown in the left panels of Figure~\ref{fig:plstoy}, as expected. This is because $\Xb_1$ only had an object mean added to its shared signal. However, the first $\Xb_2$ component is completely dominated by the large linear trend in the trait mean rather than one of the underlying shared effects. This is a consequence of PLS choosing score vectors to maximize covariance rather than correlation. The trait mean effect is much larger in magnitude than the underlying shared structure, so the best way to maximize covariance is to choose a score vector close to the trait mean effect for $\Xb_2$.

\begin{figure}[H]
\includegraphics[width=\textwidth]{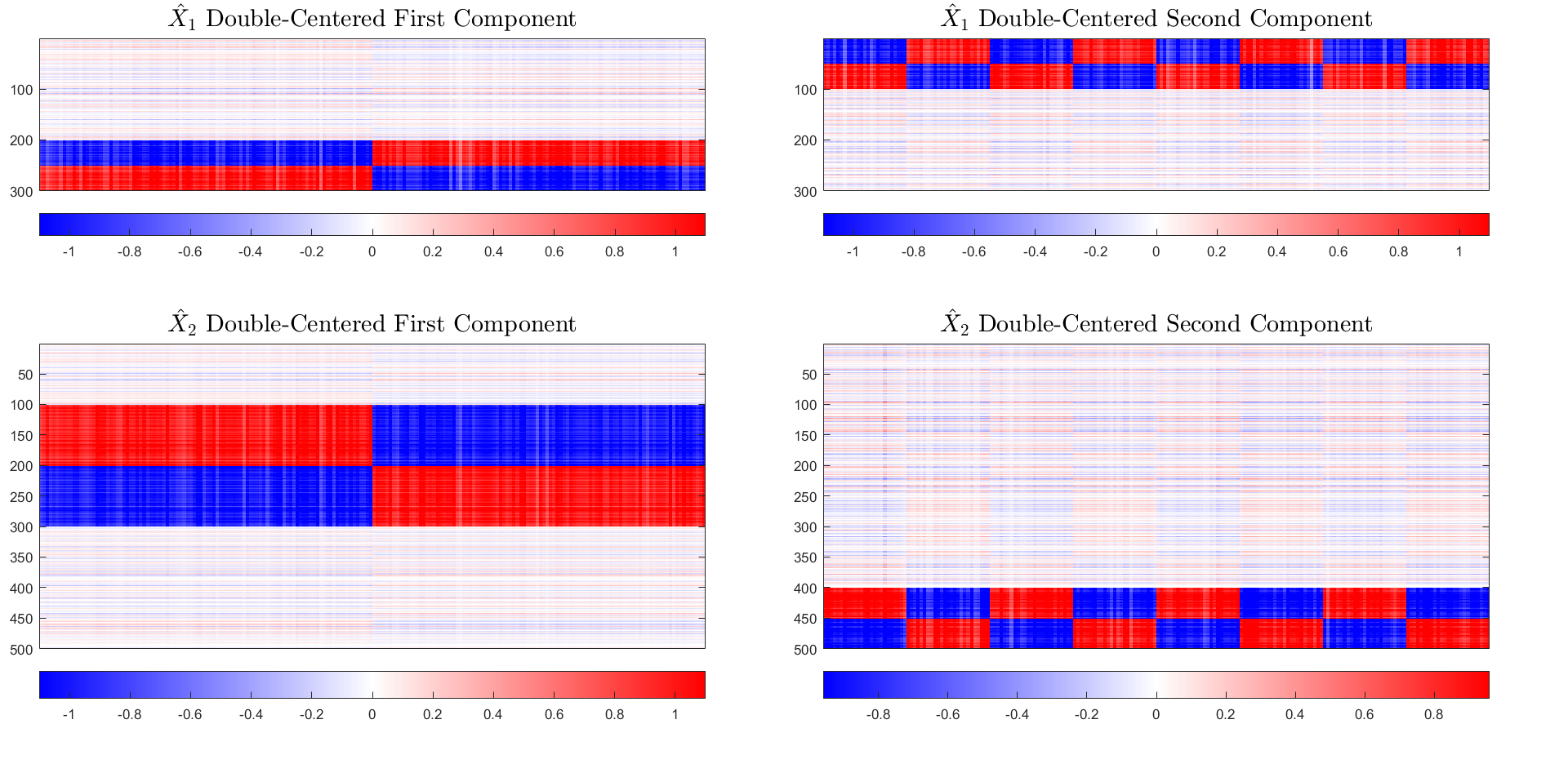}
\caption{First two PLS components of each double centered synthetic data block. Recovery and parsing of distinct underlying signal pieces is strong for both blocks.}\label{fig:plsdbl}
\end{figure}

Figure~\ref{fig:plsdbl} displays the first two PLS components of each data block found using the double centered versions of the matrices. The top panels show the $\Xb_1$ components and the bottom two panels show the $\Xb_2$ components. Now that the strong trait mean effect in $\Xb_2$ has been removed, the recovery of the underlying shared signal is greatly improved in both blocks. The first component in both blocks is distinctly the long-checkered pattern and the second component in both blocks is distinctly the short-checkered pattern.

In this synthetically constructed example, a strong trait mean effect dominated the calculated components from data integration. Removal of the trait means of each block in addition to the typical and necessary removal of the object means of each block drastically improved interpretability of results.

\subsection{Spanish Mortality: Males versus Females}

We return to the Spanish mortality data from Section~\ref{sec:fda} to further explore the implications of additional centering in data integration tasks. Here we combine the observations of male mortality rates from 1908 to 2002 with corresponding measurements of female mortality rates over the same time period. We perform PLS on these two data blocks to locate the shared information between them. We opt for the algorithmic approach outlined in Section~\ref{subsec:pls} to ensure score vectors are orthogonal. We will compare the analysis after object centering and double centering. In both of these centering regimes, the score vectors will be mutually uncorrelated (See Table~\ref{tbl:uncorr}).

\begin{figure}[H]
\centering
\includegraphics[width=\textwidth]{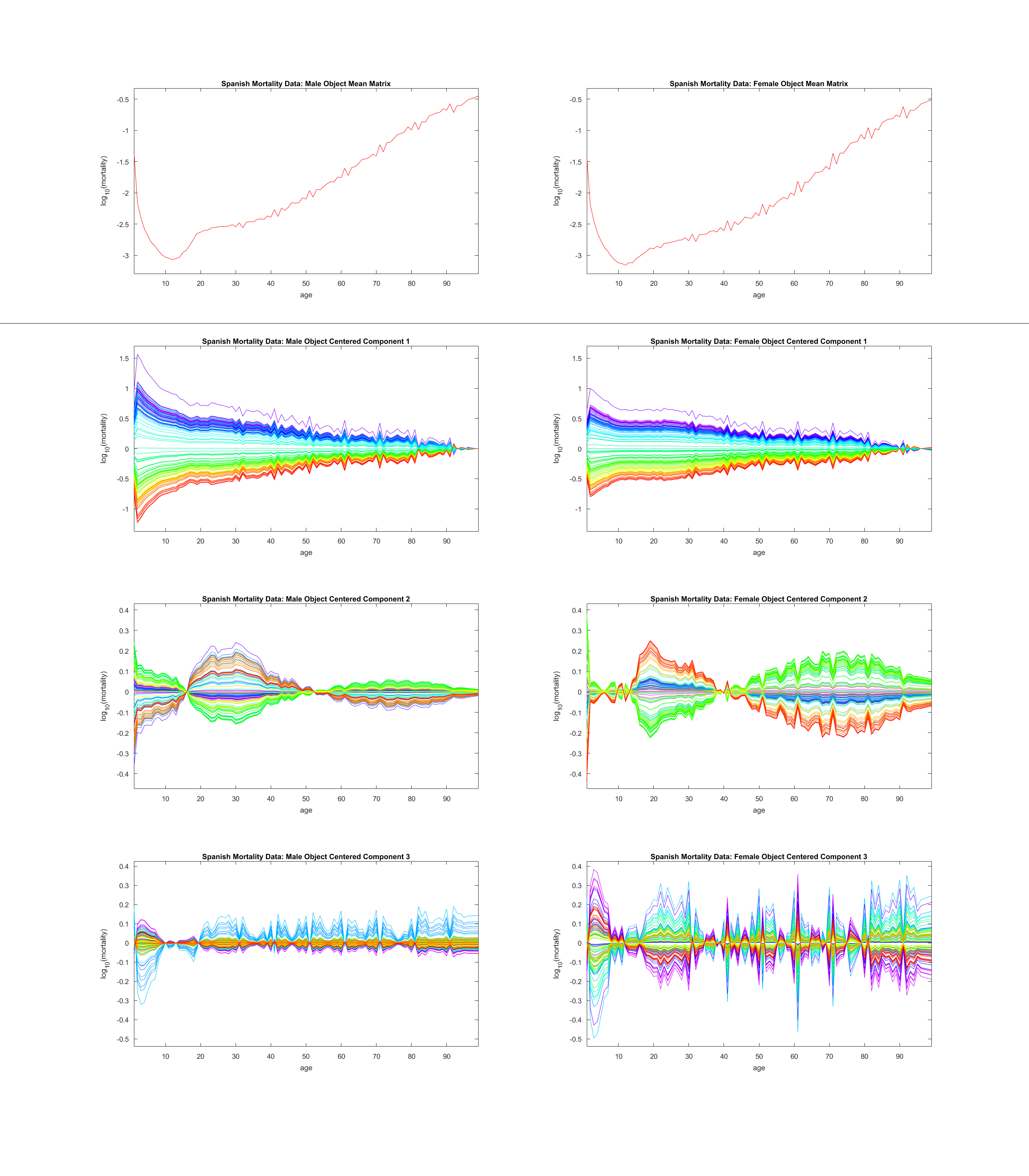}
\caption{Object-centered PLS of male and female mortality rates. First and second mode show expected trends, but third mode is challenging to interpret.}\label{fig:mortplsobjectcurves}
\end{figure}

Figure~\ref{fig:mortplsobjectcurves} shows the results of PLS on the two data blocks after each has been object centered. We display the loadings vectors scaled by the scores of each observation in a similar fashion to Figures~\ref{fig:mortobjectcurves}, \ref{fig:mortnoncurves}, and \ref{fig:mortdoublecurves}. The object mean and first three joint modes of variation for males are shown on the left, and the corresponding modes for females are shown on the right. While each mode of variation manifests differently in each data block, the same broad trends are identifiable for each gender. The first mode shows overall improvement and more dramatic improvement for younger people, and the second mode shows a contrast between younger adults and the rest of the population. This contrast highlights ages 18-50 for males and ages 15-45 for females. The third mode is much harder to interpret as there is no obvious commonality between the patterns for each gender outside of the appearance of age rounding. Overall these modes correspond with those found via the PCA analysis of male mortality in Section~\ref{sec:fda}. Since PCA finds modes of maximal variation and PLS tries to find directions with maximal covariance between blocks, this correspondence between PCA and PLS modes of variation is not surprising.


\begin{figure}[H]
\centering
\includegraphics[width=\textwidth]{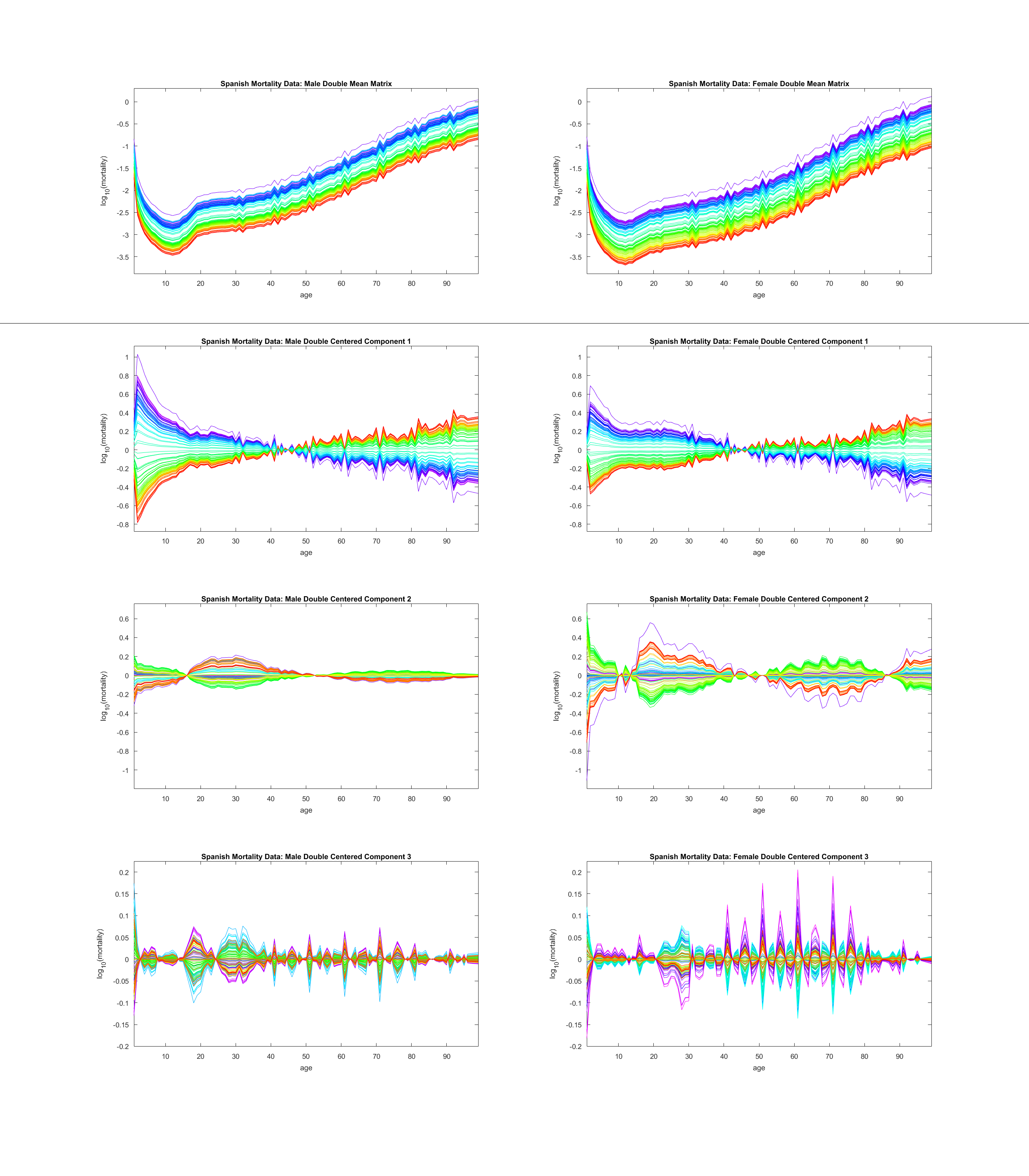}
\caption{Double-centered PLS of male and female mortality rates. First and second mode show expected trends, and third mode highlights record-keeping anomalies in each gender.}\label{fig:mortplsdoublecurves}
\end{figure}

Figure~\ref{fig:mortplsdoublecurves} shows the results of PLS on the two data blocks after each has been double centered, organized in a similar fashion to Figure~\ref{fig:mortplsobjectcurves}. Again the first two modes of variation match expectations. The first mode shows stronger improvement for younger individuals as the overall improvement has been removed with the trait mean, and the second mode again shows a contrast between younger adults and the rest of the population. The third mode now more clearly pertains to age rounding for both males and females. In addition to large spikes every ten years, we also see smaller spikes every five years, further reflecting a bias towards rounder numbers on death certificates of older individuals.

As in previous analyses of this kind of data, we feel the choice to double-center each data block enhances the interpretability of the results.

\section{Conclusions}

In this manuscript we presented a unified framework for describing, understanding, and implementing centering as part of data matrix analysis. We put forth disambiguating terminology for describing data matrix dimensions and centering operations. We highlighted double centering in FDA as a way to incorporate the constant function direction mode of variation in data analyses. Correspondingly, we proposed a hypothesis test for determining whether the constant function direction mode of variation is significant for a given data set.

Finally, we explored the implications of different centerings in multi-block data integration settings. While we focused on data with two blocks of traits pertaining to the same data objects, many modern data integration tasks involve large combinations of blocks that can share either data objects or traits in common. Potential centerings in both object space and trait space must be fully explored and understood as part of these analyses, and our framework provides the necessary tools for that exploration and understanding. 


%
%

 \section*{Acknowledgements}
 
The authors would like to thank Charles Perou, Katherine Hoadley, Chris Fan, and Andres Alonso for the collection and maintenance of the data used in this manuscript. This research was partially supported by the National Science Foundation under Grant No. IIS-1633074 and DMS-1916115.
 
%
%



\bibliographystyle{imsart-nameyear} 
\bibliography{proposalRef}       


\end{document}